\documentclass[11pt]{article}

\usepackage{geometry}
\usepackage[utf8]{inputenc}
\usepackage{amsmath}
\usepackage{subcaption}
\usepackage[font=small,skip=1pt]{caption}
\usepackage[numbers]{natbib}
\usepackage{tikz}
\usepackage{float}
\usepackage{multirow}
\usetikzlibrary{patterns}

\usepackage{pgfplots}
\pgfplotsset{width=10cm,compat=1.9}

\usepackage{authblk}

\usepackage{amssymb, amsthm, amsmath, amsfonts, mathrsfs} %
\usepackage{comment}

\usepackage{color, soul} 
\usepackage{xcolor}  

\newcommand{\gray}{\textcolor{\gray}}

\newcommand{\black}{\textcolor{black}}

\usepackage{hyperref} 
\hypersetup{
	colorlinks=true,
	citecolor={blue}, 
	urlcolor={blue}, 
	linkcolor={blue}, 
	breaklinks={true}
}

\title{\textbf{On the optimal layout of a dining room in the era of COVID-19 using mathematical optimization}}

\author[$(1)$]{Claudio Contardo \thanks{C. Contardo: \url{https://orcid.org/0000-0001-7595-3904} / claudio.contardo@gerad.ca}}
\affil[$(1)$]{\small ESG UQAM and GERAD, Montreal, Canada.}

\author[$(2)$]{Luciano Costa \thanks{L. Costa: \url{https://orcid.org/0000-0002-6324-6556} / luciano.ccosta@ufpe.br}}
\affil[$(2)$]{\small  Tecnology Department, Federal University of Pernambuco, Caruaru, Brazil.}

\date{}

\usepackage{fancyhdr}
\begin{document}

\maketitle

\begin{abstract}
We consider the problem of maximizing the number of people that a dining room can accommodate provided that the chairs belonging to different tables are socially distant. We introduce an optimization model that incorporates several characteristics of the problem, namely: the type and size of surface of the dining room, the shapes and sizes of the tables, the positions of the chairs, the sitting sense of the customers, and the possibility of adding space separators to increase the capacity. We propose a simple, yet general, set-packing formulation for the problem. We investigate the efficiency of space separators and the impact of considering the sitting sense of customers in the room capacity. We also perform an algorithmic analysis of the model, and assess its scalability to the problem size, the presence of (or lack thereof) room separators, and the consideration of the sitting sense of customers. We also propose two constructive heuristics capable of coping with large problem instances otherwise intractable for the optimization model.
\end{abstract}

\section{Introduction}\label{sec:introduction}

The current pandemic has posed major challenges to multiple economic sectors. In particular, the tourism sector is one of the most affected by the multiple restrictions imposed by health organizations and governments aiming at protecting the public from a rapid and uncontrolled spread of the disease known as COVID-19 \citep{Dube2020}. Vaccination and social distancing are perhaps the two most efficient mechanisms to reduce the propagation of the disease. In the past few months, a global vaccination campaign has proven efficient at reducing its spread, but remains insufficient as an isolated mechanism to mitigate its impact.

\begin{figure}[!ht]
	\centering
	\includegraphics[trim=7.5cm 0.35cm 7.5cm 2.5cm, clip=true, scale=0.185]{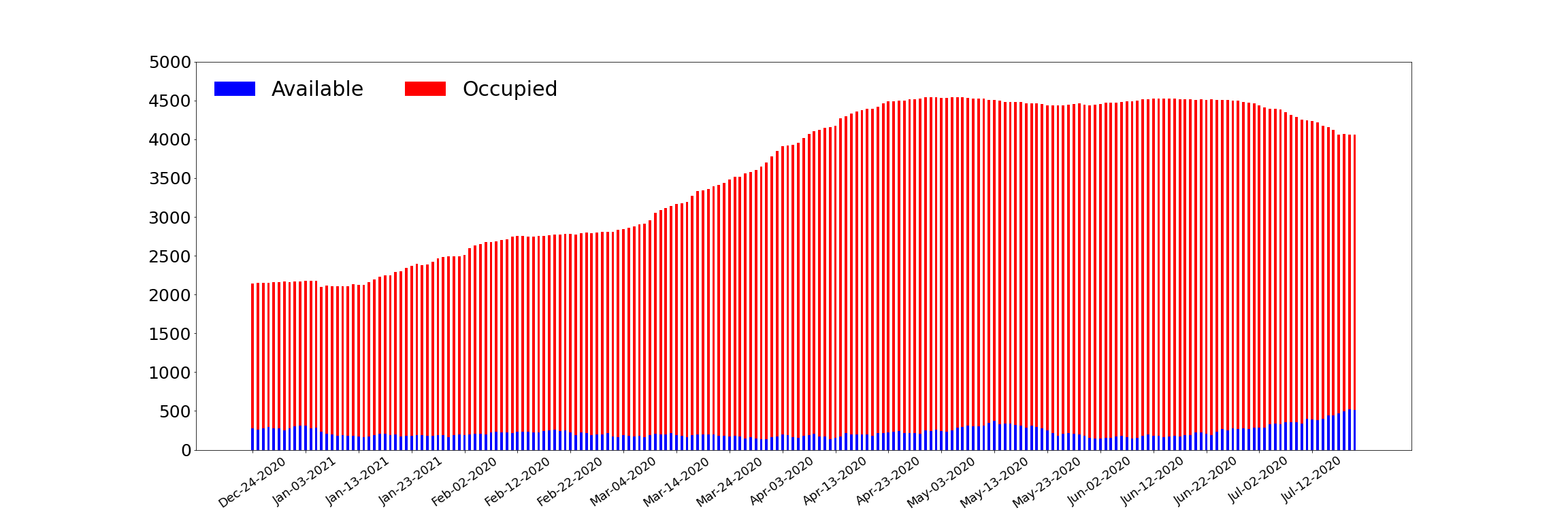}
	\caption{Comparison between the number of available and occupied hospital beds in Chile}
	\label{fig:bed_occup}
\end{figure}

\begin{figure}[!ht]
	\centering
	\includegraphics[trim=7.5cm 0.35cm 7.5cm 2.5cm, clip=true, scale=0.185]{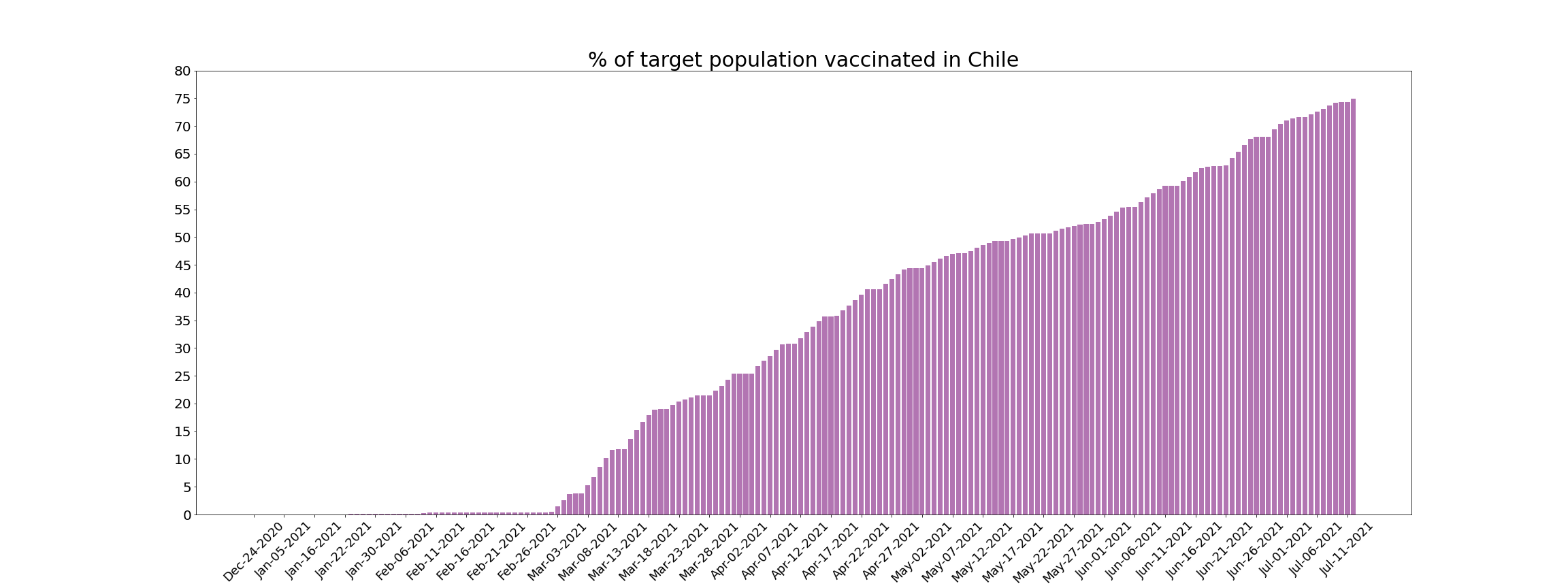}
	\caption{Percentage of the target population that is fully vaccinated in Chile}
	\label{fig:vaccionation_ap}
\end{figure}

\black{Chile is a remarkable example of such attempt. According to data from the \citet{MHC2021}, even with vaccinations attaining global maximums in July/2021 (almost 75\% of the population is fully vaccinated, meaning that they received either a single-dose vaccine, or the two doses of a two-dose vaccine), the spread of the disease did not decrease at the same pace. The rapid relaxation of social distancing measures by the local health authorities has been named as one of the main factors to explain a continuing increase in hospitalizations (Figure \ref{fig:bed_occup}), despite the success of the vaccination campaign (Figure \ref{fig:vaccionation_ap}). In certain countries, the lack of a strong vaccination campaign or the resistance of the population to get vaccinated leaves social distancing as the main mechanism to control the spread of the disease.}

Additionally, it has become apparent that the environmental conditions play a role in defining social distancing rules. Cold and temperate climates are more favorable to the spread of the virus, as compared to warmer climates \citep{araujo2020spread}. Outdoors gatherings are also less favorable for the spread capabilities of the virus as compared to indoors \citep{freeman2020covid}. In this context, the presence or absence of wind is also believed to be relevant in determining the spread capabilities of the virus \citep{feng2020influence}. In closed areas, the ventilation capabilities of rooms seem to also correlate negatively with the spread capabilities of the virus \citep{Dbouk2020, somsen2020small}.

\begin{figure}[!ht]
	\centering
	\includegraphics[scale=0.5]{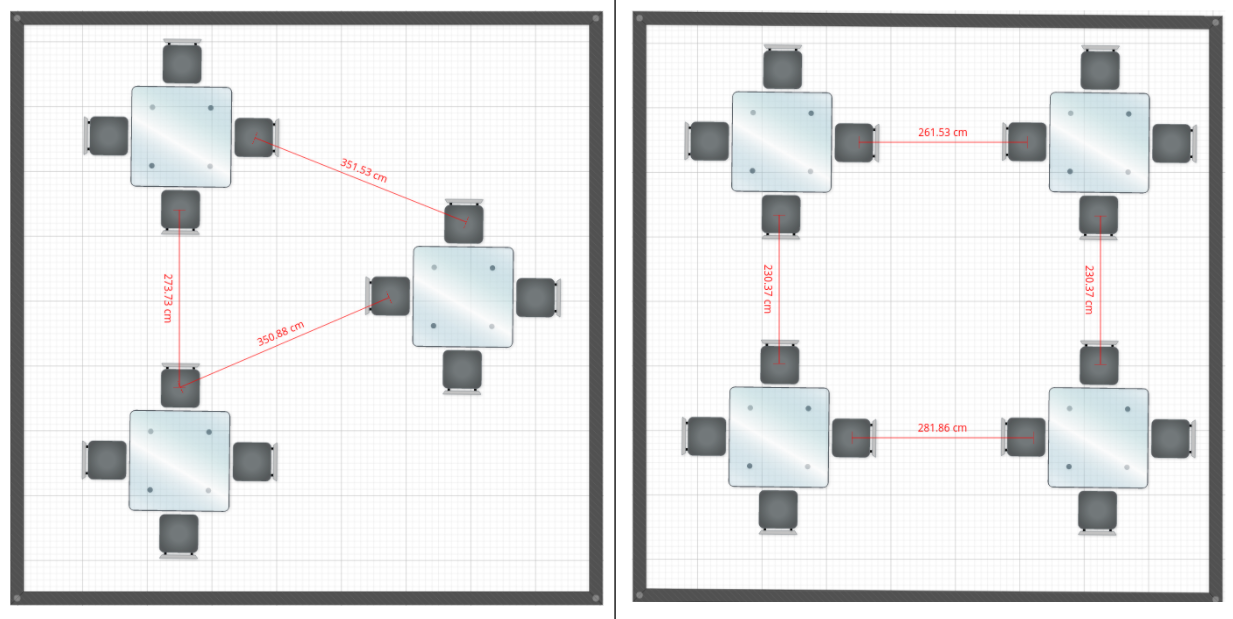}
	\caption{Example of tables layout}
	\label{fig:example_layout_tables}
\end{figure}

In the presence of constraints arising from social distancing, finding the best disposition for tables in the dining room of a restaurant can be a challenging task. In trying to satisfy social distancing, restaurants may end by not efficiently using part of the available dining room area. Think for instance of the case where a decision maker imposes a distance of two meters or more between any two chairs belonging to different tables. Figure \ref{fig:example_layout_tables} provides two examples of layouts over the same area for which social distancing using this rule is respected, but for which in one of them (left layout), the available area is not efficiently occupied. 

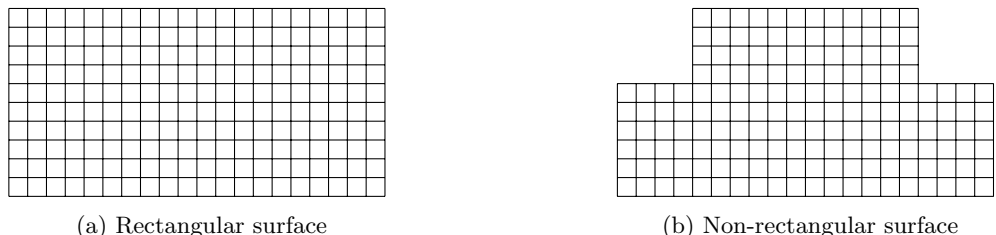
\begin{figure}[!ht]
	\centering
	\begin{subfigure}{0.49\textwidth}
		\centering
		\begin{tikzpicture}[scale = 0.5]
\draw[step=0.5,black,thin] (0,0) grid (10,5);
\end{tikzpicture}
		\caption{Rectangular surface}
		\label{fig:rectangular}
	\end{subfigure}%
	\begin{subfigure}{0.49\textwidth}
		\centering
			\begin{tikzpicture}[scale = 0.5]
	\draw[step=0.5,black,thin] (0,0) grid (10,3);
	\draw[step=0.5,black,thin] (2,3) grid (8,5);
	\draw (2, 3) to (2, 5);
	\end{tikzpicture}
		\caption{Non-rectangular surface}
		\label{fig:nonrectangular}
	\end{subfigure}
	\caption{Examples of surfaces}
	\label{fig:example_surfaces}
\end{figure}

In a small example, like the one presented in Figure \ref{fig:example_layout_tables}, at a first glance, locating tables to sit as many people as possible in the space available may seem easy. However, when we consider larger areas, which might not always be defined over rectangular surfaces (Figure \ref{fig:example_surfaces}), and that sometimes may present some physical obstacles (e.g. stairs and/or support columns), the task of positioning tables adequately may become very difficult.

In the case where dining rooms layouts do not use all the available space, businesses may experience a drastic reduction in their profit margins, which may cause them to possibly even going bankrupt \citep{Dube2020}.
\black{At the same time, restaurant owners cannot simply disregard safety measures in an attempt to increase dining rooms capacities, as customers' perception may be negatively impacted by crowded spaces, which may affect customers' choice behavior \citep{Park2021}.}
Therefore, being capable of efficiently positioning tables in a given area,
\black{or to adequately adopt safety measures}
is critical from a business perspective. In this context, decision support systems based on optimization techniques may arise as suitable tools to solve the problem of finding efficient layouts for restaurant dining rooms, while still satisfying social distancing constraints. Optimization methods are capable of efficiently tackling complex problems that are subject to different types of requirements such as the ones emerging in times of COVID-19. As a matter of fact, recently, several optimization approaches have been developed to help to cope with some of the new challenges imposed by the pandemic.

In this paper, we address the problem of designing optimal dining room layouts using mathematical optimization. Our contribution is fourfold. First, we present a simple, yet very generic, optimization model to maximize the number of people that a dining room can accommodate. Our model aims at finding the best way to place tables in a dining room by taking into account multiple attributes relevant in practice, namely: the configuration of tables, the presence of (or lack thereof) obstacles, the shape of the dining room, and the sitting sense of the customers, all of which are fully integrated in our model. Second, we perform a thorough computational analysis to understand the impact of certain attributes ---such as the sitting sense of customers, or the installation of plexiglass divisions--- in the room capacity. Third, we propose two constructive heuristics that are capable of dealing with large instances of the problem in reasonable computing times, and that for scenarios where there are no obstacles in the room, can find (near-)optimal solutions. Fourth, in an attempt to contribute to the tourism sector in the process of having to accommodate their dining rooms layouts during the current pandemic, we provide an open-source Julia package for public download and use.

The remainder of the paper is structured as follows. In Section \ref{sec:litrev}, we review the scientific literature on the use of quantitative tools ---with a particular emphasis on optimization--- during the current pandemic.
In Section \ref{sec:proposed_model}, we provide a detailed description of the problem and present the proposed formulation.
In Section \ref{sec:computational_experiments}, we generate instances for the problem, analyze the scalability of our model, describe the proposed heuristics, and provide some managerial insights.
Finally, in Section \ref{sec:conclusions}, we present some final remarks and draw some conclusions regarding the use of our approach in the context of COVID-19.

\section{Literature review\label{sec:litrev}}

In this section we review the scientific literature on recent attempts to use optimization and other quantitative tools to help decision makers make better decisions during the current pandemic.

\citet{Duque2020} introduce a mathematical model to allow policymakers make decisions regarding when to relax and when to enforce social distancing in the city of Austin, USA. The model provides optimal thresholds that, if attained, would serve as triggers to indicate when social distancing should be relaxed or enforced. Even if social distancing is one of the most effective measures to slow down the propagation of COVID-19, it may incur high costs for the society. In this context, the authors claim that the thresholds in the model are established in such a way to minimize the number of days in which (expensive) social distancing is required, while still guaranteeing that the capacity of hospitals is not exceeded with hospitalized people.

\citet{Pacheco2020} implement a Tabu Search heuristic to solve a routing problem arising in the context of urgent delivery operations of face shields in the province of Burgos (Spain). In the considered problem, vehicles and drivers are part of different volunteer organizations. For this reason, routes may be open, i.e., they can start and end at different points. Routes in the problem include delivering raw materials to produce the masks, collecting face shields, and delivering them to the demand points. Given that drivers are volunteers, routes should be generated in such a way the working time among them is as similar possible. Hence, the objective considered in the problem was the minimization of the longest route generated. Despite simple, the proposed method contributed significantly to the combat of the pandemic in the province. The routing system deployed has allowed Burgos to have a relative number of face shields in comparison with the population of the province of 10.5\%, whereas in the rest of Spain this value was inferior to 1.5\%.

\citet{Seccia2020} and \citet{Zucchi2020} propose mathematical models to tackle the problem of finding balanced personnel schedules in contexts where COVID-19 imposes additional constraints. \citet{Seccia2020} solve a nurse rostering problem where the number of nurses is insufficient to meet the demand. To face this problem, the model allows nurses to work more than one shift per day. Yet, to reduce the psychological stress (in excess) to which nurses might be exposed, the model aims to balance the overall number of working hours assigned to all nurses. \citet{Zucchi2020} address a personnel scheduling problem motivated by a real-life application found in an Italian pharmaceutical distribution company. Before the pandemic, workers used to work in different sectors throughout the working day. However, during the pandemic, in an attempt to reduce the risk of contagion, the company has decided to define mutually exclusive groups of workers, who should always work together. These groups are defined so as the total deviation between the total contractual hours and the worked hours is the minimum possible.

\textcolor{black}{\citet{Ugail2021} propose a decision-support tool to enable the optimal configuration of a given space in the presence of social distancing requirements. The authors model the problem of locating a series of points in a two-dimensional region as a circle-packing problem \citep{Dubejko199519}. The model is specific to the region under consideration, leading to different models depending if the region is circular, triangular, or rectangular, for instance, and therefore less susceptible to generalizations to arbitrary areas. The authors show that their models can handle air flows as the ones induced by ventilation systems with little effort, as well as the positions of windows and doors.}

\citet{Fischetti2020} present a mathematical model for the problem of fitting facilities (e.g., customers and beach umbrellas) into a given area while respecting social distancing constraints. The authors adopt a formulation initially proposed for farm layout optimization problems, in relying on the similarity between both problems. The proposed model selects from a set of predefined positions the ones where facilities should be placed while ensuring a minimum distance between the facilities and minimizing infection risk. In the study, the infection risk is assessed through different exposure measures which are a function of the Euclidean distance. When considering two types of areas -- 1- and 2-dimensional areas --, exposure indicators defined as being inversely proportional to the squared or the cubic power of the Euclidean distance yielded the most satisfactory layouts. Despite representing a good alternative to tackle the problem of locating facilities in a given area, the proposed methodology has two limitations that may result in solutions that do not use all the available area. First, in the approach, the maximum number of facilities is a parameter of the model, which may not be the largest possible, and hence may entail sub-optimal solutions. Second, the method does not consider topology and physical characteristics of the different components of the problem (tables sizes, sitting positions, sitting sense, surface topology). This aspect may be very important in situations in which the available space is small, such as a restaurant dining room, and we intend to use the available area to its extent.

\citet{Wang2021} employ a DAST framework \citep{Roggeveen2020} to assess how environmental factors -- crowdedness (numbers of persons in a service environment) and restaurant safety measures -- affect costumers consume choices and their perceptions during the COVID-19 pandemic. In the study, USA (country highly impacted by the COVID-19) and Australian (country known to have effectively controlled the spread of the disease) residents were interviewed. The results have shown that US residents are more sensitive to crowdedness, even moderate. According to the study, US residents are less likely to eat in restaurants than Australians during the pandemic. This is probably due to the low perceived severity of the pandemic in Australia. Regarding restaurant safety measures, both Americans and Australians seem to prefer social distancing over the use of partition screens. The latter may require an initial investment, thus, for the restaurant, it may be perceived as more expensive. Moreover, this measure may reduce the flexibility of changing the setting of tables and the restaurant layout. Yet, this approach allows restaurants to increase the capacity of their dining rooms.

Even if they were not (initially) motivated by room layout problems, some research works that might be applied in a context similar to the one discussed in this paper are those related to two-dimensional cutting and packing problems (2D-CPPs). In 2D-CPPs, items and bins that may have regular (e.g., rectangles) or non-regular (convex or non-convex) shapes must be packed. Hence, 2D-CPPs have some similarities with the problem description presented in Section \ref{sec:introduction}. A natural geometrical problem arising in the context of the 2D-CPPs is due to non-overlapping constraints. Several approaches have been proposed in the literature to address this issue. \citet{Sato2013} propose three placement heuristics for irregular 2D-CPPs, all of which relying on the notion of the collision-free region and on its ability to define exact fitting and exact sliding placements. Another technique largely employed to deal with the overlapping issue is the consideration of a set of points or patterns. In this technique, the position of an item into a bin is identified by the location of its bottom-left corner that is represented by a point $(x,y)$. This representation has been considered by \citet{Silva2016}, who present a review on solution methods for the pallet loading problem, a variant of 2D-CPP that aims to pack rectangular boxes onto a single rectangular pallet, with orthogonal rotation being allowed. Later, \citet{Cote2018} propose a new set of patterns called meet-in-the-middle, which are a combination of two types of patterns. Meet-in-the-middle patterns possess interesting properties that yield a considerable reduction in the cardinality of the set of points when compared to the traditional approach.

\citet{Bezerra2020} develop several formulations for the strip packing problem, a variant of 2D-CPP, where rectangular items must be packed into a rectangular area (strip) in such a way there is no overlapping between items, and the height of the strip is maximized. The proposed formulations rely on different paradigms to deal with overlapping, namely, packing in levels and the consideration of a graph on unoccupied positions. \citet{AlvarezValdes2018} give a general overview about 2D-CPPs, where the authors discuss aspects such as dimensionality, geometry, as well as additional characteristics and constraints defining different variants of 2D-CPPs. Besides, a broad overview of the main heuristic techniques used to tackle 2D-CPPs is also provided. Finally, a recent survey on exact solution methods to tackle 2D-CPPs is presented by \citet{Iori2021}.

\section{The proposed model}\label{sec:proposed_model}

In this section, we describe how we model the problem of finding optimal layouts for dining rooms as a set-packing problem while still assuring that social distancing constraints are satisfied.
This section is structured as follows. In Sub-section \ref{sec:roomtop}, we describe our modeling of the room topology, including several features that are relevant in practice, such as the existence of (or lack thereof) obstacles. 
In Sub-section \ref{sec:definition_sitting_configuration}, we detail how we encode into the variable definition of our formulation the several characteristics (topology and physical aspects) associated with the surface over which we intend to determine the optimal layout.
In Sub-section \ref{sec:conflict_graph}, we present how we impose social distance requirements in our model.
Finally, in Sub-section \ref{sec:formulation}, we present the set-packing formulation that we adopt to solve our layout problem.

\subsection{Room topology}\label{sec:roomtop}

\begin{figure}[!ht]
		\centering
			\begin{tikzpicture}[scale = 0.8]
	\draw[step=0.5,black,thin] (0,0) grid (10,3);
	\draw[step=0.5,black,thin] (2,3) grid (8,5);
	\draw[fill=gray] (0.1,0.1) rectangle (1.9,0.9);
	\draw[fill=gray] (2.1,2.1) rectangle (2.4,2.4);
	\draw[fill=gray] (7.1,3.1) rectangle (7.4,3.4);
	\draw[fill=gray] (8.1,0.1) rectangle (8.4,0.9);
	\draw[pattern = north west lines, pattern color = gray] (4.2, 1.25) rectangle (4.3, 3.25);
	\draw[pattern = north west lines, pattern color = gray] (4.75, 3.7) rectangle (6.75, 3.8);
	\draw (2, 3) to (2, 5);
	\end{tikzpicture}
		\caption{Surface with the representation of physical obstacles}
		\label{fig:topology_physical_characteristics}
	\end{figure}
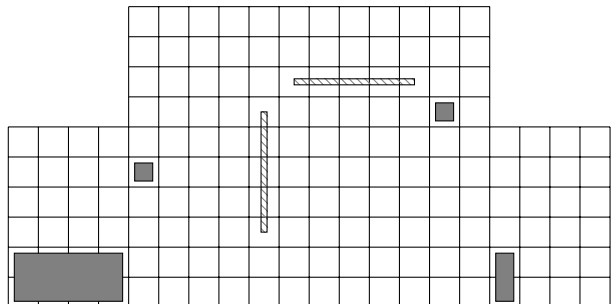

We consider a discretization of a physical space in the form of a grid, which is not required to be rectangular. Each block in the grid represents a position in the surface where we may (or not) locate a table. Some blocks in the grid may be marked as \textit{soft obstacles}, this is, objects that prevent the location of tables, but do not stop the circulation of air particles through them (think of plants or flowers, for instance). Some other blocks in the grid may be marked as \textit{hard obstacles} which not only prevent the location of tables but also the passing of air particles through them (think for instance of plexiglass or other type of walls). In Figure \ref{fig:topology_physical_characteristics}, we depict an example of such grid, where soft obstacles are marked in solid gray while hard obstacles are marked as thin dashed rectangles.
	
	\subsection{Sitting configuration}\label{sec:definition_sitting_configuration}
	
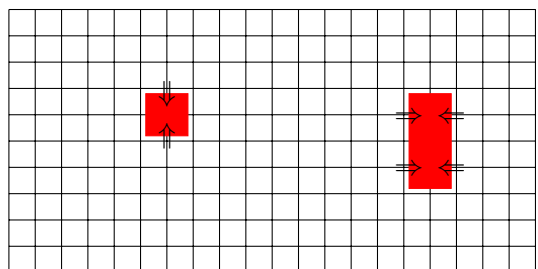
\begin{figure}[!ht]
			\centering
			\begin{tikzpicture}[scale = 0.7, every node/.style={minimum size=.5cm-\pgflinewidth, outer sep=0pt}]
	\draw[step=0.5,black,thin] (0,0) grid (10,5);
	\draw[red, fill = red] (2.6, 2.6) rectangle (3.4, 3.4);
	\node at (3, 3.4) {$\Downarrow$};
	\node at (3, 2.6) {$\Uparrow$};
	\draw[red, fill = red] (7.6, 1.6) rectangle (8.4, 3.4);
	\node at (7.6, 1.95) {$\Rightarrow$};
	\node at (7.6, 2.95) {$\Rightarrow$};
	\node at (8.4, 1.95) {$\Leftarrow$};
	\node at (8.4, 2.95) {$\Leftarrow$};
	\end{tikzpicture}
			\caption{Example of two sitting configurations}
			\label{fig:sitting_config}
		\end{figure}

	We assume that each sitting configuration uses a certain number of blocks in that grid (the finer the grid, the larger the number of blocks each sitting configuration will use), and are not allowed to overlap with any type of obstacle. In Figure \ref{fig:sitting_config}, we illustrate two possible sitting configurations: 1) a single square table capable of accommodating two persons sitting along a vertical orientation axis; 2) two tables in a row, capable of accommodating four persons, who sit along a horizontal orientation axis. Note that these two configurations are presented only for illustration purposes. The model that we will present can handle arbitrary types of tables provided that they are defined in advance. In this context, a sitting configuration is defined in terms of: i) the shape of the table (e.g., square, rectangle, etc); ii) the number and the position of the chairs in the table (which implicitly establishes the sitting orientation of the persons); and iii) the position of the sitting configuration in the room, which is given by the coordinates of the blocks of the grid that are occupied by the table.
		
		\subsection{Social distancing restrictions}\label{sec:conflict_graph}
		
		Similar to the strategy that we employ to model physical/topological aspects associated with the surface over which we solve our problem, the approach that we adopt to impose social distancing requirements in our model is also very generic. We represent distancing constraints by means of a conflict graph. The representation that the conflict graphs assume in our study is richer than that found in \citet{Fischetti2020}. In \citeauthor{Fischetti2020}'s model, an edge in the conflict graph indicates that two positions in an area are not located far enough from each other. In our case, edges in the conflict graph correspond to \textit{incompatible} sitting configurations. Two sitting configurations are said to be \textit{compatible} if: 
		\begin{enumerate}
			\item they do not overlap; \texttt{and}
			\item for every pair of chairs in different tables, either:
			\begin{enumerate}
				\item they are more than two meters (equivalently six feet) distant from each other; \texttt{or}
				\item they are separated by a hard obstacle (a wall); \texttt{or}
				\item they are oriented back to back forming an angle of 180$^{\circ}$ \citep{Dbouk2020}.
			\end{enumerate}
		\end{enumerate}
		
		In this context, a brief description of how we build a conflict graph can be given as follows. At first, we generate all the possible sitting configurations that might be placed in the room. We iterate over each block $(i, j)$ in the grid and generate up to four tables, namely: 1) one squared table located at block $(i, j)$ with the customers seated along the horizontal axis; 2) one squared table located at block $(i, j)$ and with the customers seated along the vertical axis; 3) one horizontal rectangular table occupying blocks $(i, j), (i, j + 1)$; 4) one vertical rectangular table occupying blocks $(i, j), (i + 1, j)$. We of course pay attention to not generate sitting configurations that would go out of bounds or intersect with an obstacle.
		
		As formally mentioned in Section \ref{sec:formulation}, each sitting configuration constitutes a node in the graph. In turn, one defines an edge in the graph for every pair of incompatible sitting configurations. Given two sitting configurations, we proceed as follows to assess if they are incompatible. At first, we check if the minimum distance between any two persons sitting at the table is less than 2m (6 ft). If that is not the case, we verify at which angle these two persons are sitting from each other. If they are not sitting back to back (angle of 180$^o$), we proceed to the last test that consists of verifying if there is a hard obstacle between them. To perform this verification, we check if the (imaginary) line segment connecting any two persons from different tables, and that are sitting less than 2m away from each other, intersects a hard obstacle, which is also represented by a segment line. If this condition is not satisfied, the two sitting configurations are deemed incompatible and an edge in the conflicting graph is generated.
		
		Every pair of sitting configurations that do not comply with one of these conditions is marked as incompatible, meaning that the two sitting configurations cannot appear simultaneously in a feasible solution of the problem. We point out that, even for large instances (Section \ref{sec:instances}), the computing time required to build the conflict graph is very low. More specifically, for the instances considered, the computing time did not exceed five seconds.
		
		We point out that the notion of conflict graph that we consider is not limited to the mathematical formulation that we propose. In fact, we also propose two heuristics that rely on the same notion, and that can yield good quality solutions for certain types of problems.

	\subsection{Mathematical formulation}\label{sec:formulation}
	
	We now present a mathematical model to find the layout that maximizes the number of people that the dining room can accommodate. Let $G = (V, E)$ be the conflict (undirected) graph mentioned in Section \ref{sec:conflict_graph}. The node set $V$ corresponds to all possible sitting configurations in the grid. The edge set $E$ contains all possible pairs of possible sitting configurations that are not pairwise compatible. For a given sitting configuration $u\in V$, let $z_u$ be a binary variable indicating whether the configuration $u$ is chosen, and let $s_u$ be the number of persons it may accommodate. The following set-packing problem maximizes the number of persons that may sit in the dining room:
	\begin{equation}
		\max\qquad \sum_{u\in V} s_u z_u \label{eq:objfunction}
	\end{equation}
	subject to
	\begin{align}
		& z_u + z_v \leq 1 & \{u, v\} \in E \label{eq:pairwise}\\
		& z_u \in \{0, 1\} & u \in V \label{eq:variables}
	\end{align}
	The objective function \eqref{eq:objfunction} aims at maximizing the total number of persons sitting in the dining room. Constraints \eqref{eq:pairwise} are pairwise conflict constraints enforcing that two incompatible sitting configurations cannot be selected simultaneously in the solution. Finally, constraints \eqref{eq:variables} define the domain of the variables.

\section{Computational experiments}\label{sec:computational_experiments}

In this section we present a computational campaign aiming at: 1) analyzing the impact of two attributes in the room capacity; and 2) analyzing the scalability of the model to some attributes and parameters of the problem. In Sub-section \ref{sec:compusetting}, we describe the computational environment used throughout our campaign. In Sub-section \ref{sec:instances}, we describe the problem instances used in our experiments. In Sub-section \ref{sec:model_variants}, we describe four variants of our model that will help at assessing its performance on a variety of scenarios. In Sub-section \ref{sec:managerial}, we analyze the effect on the room capacity of i) installing plexiglasss divisions, and ii) considering the sitting sense of customers in the distancing requirement. In Sub-section \ref{sec:sensitivity}, we present an algorithmic analysis of the scalability of the proposed model. Finally, in Sub-section \ref{sec:heuristics}, we describe two constructive heuristics to tackle the problem, and compare their performance against the exact method (solving formulation \eqref{eq:objfunction}--\eqref{eq:variables}).

	\subsection{Computational environment}\label{sec:compusetting}
	
	We have implemented the proposed mathematical model and the two heuristics using Julia 1.6.5 \citep{Bezanson2017}, with JuMP as modeling language \citep{Dunning2017}. The implemented code has been made available at \url{https://github.com/lucianoccosta/OptimalDiningLayout.jl}. We use CPLEX 12.10 as general-purpose MILP solver. Our experiments have been executed on an Intel Xeon E5-2637 v2 @ 3.50 GHz with 128 GB of RAM.
	
	\subsection{Proposed instances}\label{sec:instances}
	
	To assess the applicability and the performance of our model, we propose grid instances with rectangular-shaped areas like the one depicted in Figure \ref{fig:rectangular}. We point out that our approach can also handle non-rectangular-shaped areas such as the ones in Figure \ref{fig:nonrectangular}. However, since the conclusions drawn from some preliminary analyses are similar to those obtained for rectangular-shaped areas, we limit our experiments to the solution of rectangular-shaped problems only.
	
	As described in Section \ref{sec:roomtop}, we consider a discretization of the physical space. Hence, each instance is associated with one grid. Each grid is composed of squares of 0.70 m $\times$ 0.70 m. The choice for this value is due to the fact that 0.70 m is the typical size of the side of a restaurant table. Hence, since we assume that tables occupy blocks of the grid (Section \ref{sec:definition_sitting_configuration}), we presume that a square table (to sit two persons) occupies one square of the grid, whereas a rectangular table (to sit four persons) occupies two squares (Figure \ref{fig:sitting_config}). We generate eight classes of square-shaped instances $n \times n$, which differ by the number $n$ of squares on each side of the grid. The number of squares considered in the instances are $n \in \{5, \, 10, \, 15, \, 20, \, 25, \, 30 ,\, 35, \, 40\}$. Each instance is denoted \texttt{n\_n\_nowalls}.
	
\begin{figure}[!ht]
		\centering
		\begin{subfigure}{0.5\textwidth}
			\centering
			\includegraphics[trim=2.0cm 1.0cm 0.0cm 1.0cm, clip=true, scale=0.055]{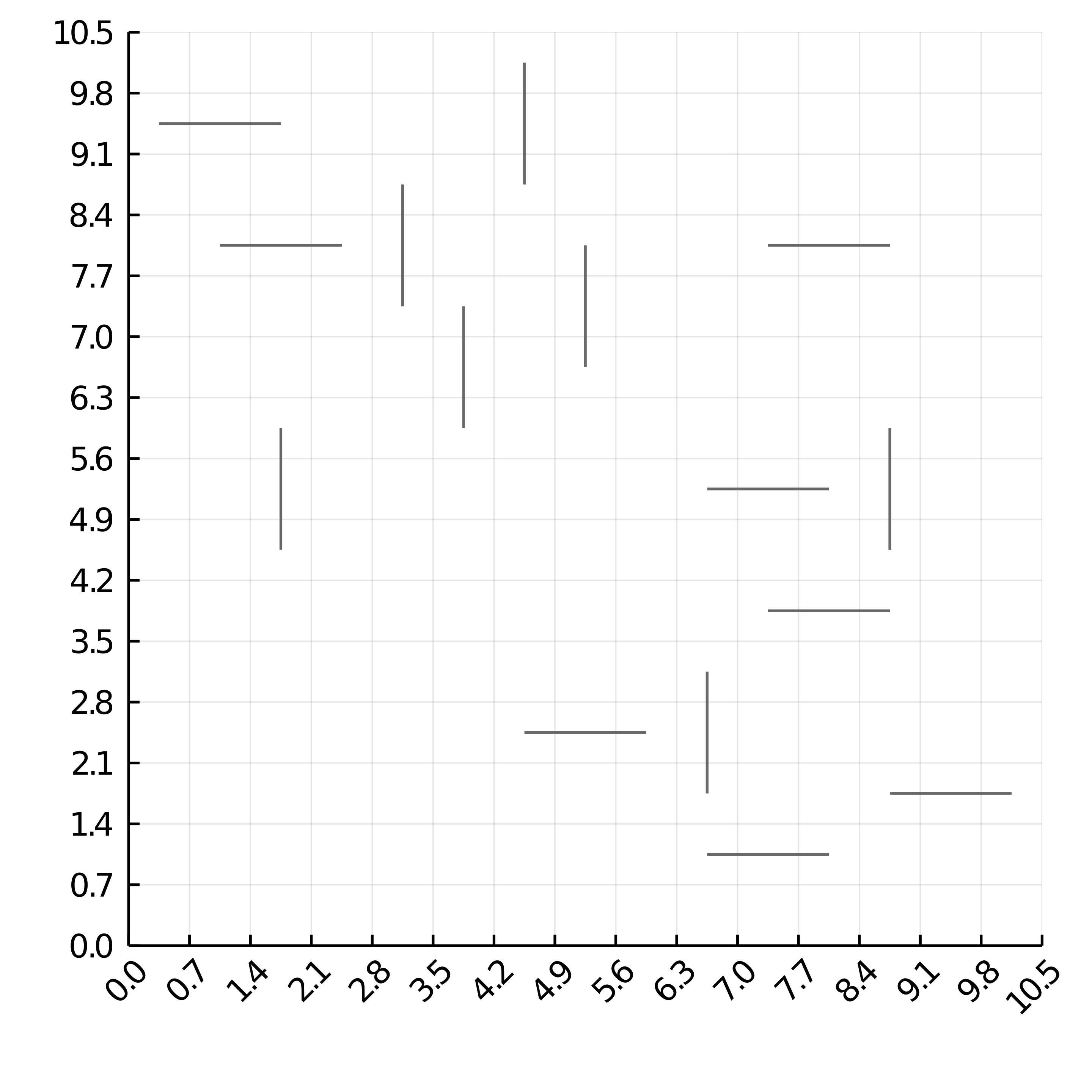}
			\caption{Random placement}
			\label{fig:instances_walls_at_random}
		\end{subfigure}\begin{subfigure}{0.5\textwidth}
			\centering
			\includegraphics[trim=2.0cm 1.0cm 0.0cm 1.0cm, clip=true, scale=0.055]{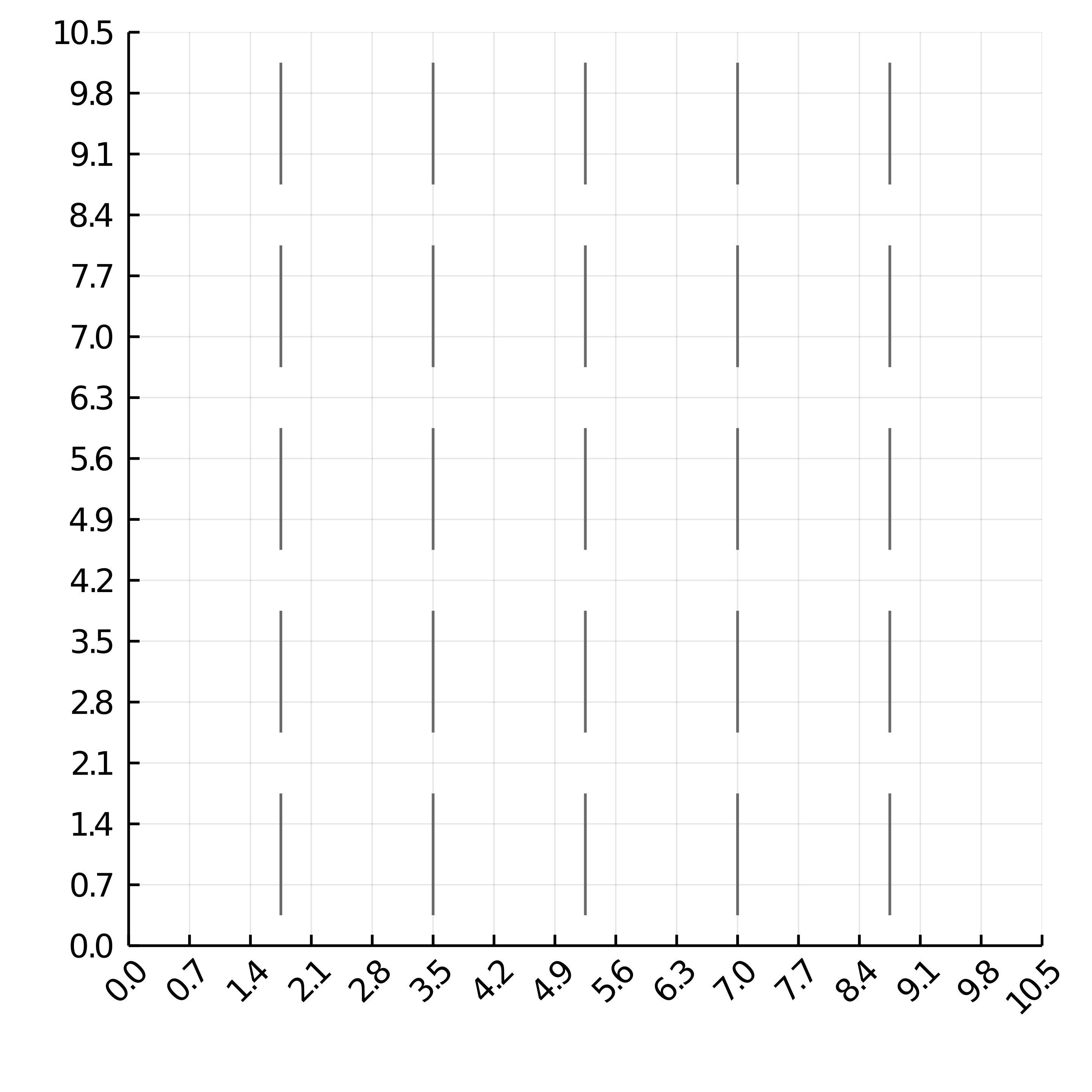}
			\caption{Uniform placement}
			\label{fig:instances_walls_uniformly_placed}
		\end{subfigure}
		\caption{Examples of random and uniform plexiglass divisions\label{fig:instances_walls}}
	\end{figure}

	In our experiments, we assess the effect on the room capacity of installing plexiglass walls (hard obstacles). To establish the location of the walls in the room, we consider two approaches: 1) locating the walls horizontally/vertically at random positions in the room (Figure \ref{fig:instances_walls_at_random}) and 2) distributing the walls uniformly vertically (equivalent to locating them horizontally) throughout the room (Figure \ref{fig:instances_walls_uniformly_placed}).
	To generate instances with plexiglass walls located at random, we proceed as follows.
	For each instance class $n \times n$, with $n \in \{10, 15, 20, 25, 30,\, 35, \, 40\}$, 	we generate $w \in \{5, 10, 15\}$ plexiglass, with lengths $0.7t$, $t \in \{1, 2, 3, 4\}$. Hence, by combining each possible value of $w$ and $t$, for each instance class, we can generate up to 12 different settings. To take advantage of the randomness of our process, for each setting, we generate 100 instances. Therefore, in total, we generate 1,200 instances. Each instance generated is denoted by \texttt{n\_n\_t\_w\_id}, with \texttt{id} being the index of the instance. To decide where to locate a plexiglass wall, at first we randomly generate a point $(x_1, y_1)$ located at the midpoint of one of the sides of a square in the grid. Then, to generate the point where the other extremity of the plexiglass wall will be placed, we select an angle $\theta \in \{0, 0.5\pi, \pi, 1.5\pi\}$ at random, and make $(x_2, y_2) = (x_1 + 0.7t \, \cos \theta, \,y_1 + 0.7t  \, \sin \theta)$. If $(x_2, y_2)$ is placed within the bounds of the dining room, we keep the plexiglass wall, otherwise, we reject it and restart the process.
	Moreover, we impose parallel plexiglass walls, intersecting in the $x$- or $y$- axis, to be placed at a distance of at least 1.4m from each other. This ensures that there will always be enough space for a table to placed between two parallel plexiglass walls, and that there will space for circulation in the room, which are reasonable assumptions.
	We repeat the process until $w$ plexiglass walls have been generated.
	
	Finally, we describe the procedure that we adopt to generate instances where plexiglass walls are uniformly distributed throughout the room. Here, we do not control the number of walls that are placed. For each instance class $n \times n$, with
	$n \in \{10, 15, 20, 25, 30,\, 35, \, 40\}$,
	and lengths $0.7t$, $t \in \{1, 2, 3, 4\}$, we try to place as many walls as possible, provided that both extremities of the segment corresponding to the plexiglass wall are placed within the boundaries of the room, and that two segments are not placed in neighbor grid blocks, otherwise, since tables are only placed in the blocks of the grid, it would not be possible to place any table between two plexiglass walls. Moreover, as depicted in Figure \ref{fig:instances_walls_uniformly_placed}, we leave an space of 0.7m (enough to place a table) between two rows of walls. As described previously, we generate instances where plexiglass walls are placed leaving a room of 0.35m from the wall (Figure \ref{fig:instances_walls_uniformly_placed}). These problems are denoted \texttt{n\_n\_uni\_t}.
	
	\subsection{Model variants\label{sec:model_variants}}
	
	Throughout or computational campaign and to support the different analysis regarding the efficiency of our modeling approach, we consider the following four settings for our model:
	\begin{description}
		\item[\texttt{Baseline}:] no plexiglass divisions are used, and the sitting sense of the customers is ignored to relax the distancing requirements,
		\item[\texttt{Plexiglass}:] plexiglass divisions are used, and the sitting sense of the customers is ignored to relax the distancing requirements,
		\item[\texttt{SittingSense}:] no plexiglass divisions are used, and the sitting sense of the customers is considered to relax the distancing requirements.
		\item[\texttt{Plexiglass+SittingSense}:] plexiglass divisions are used, and the sitting sense of the customers is considered to relax the distancing requirements.
	\end{description}
	
	\subsection{Two attributes that affect the room capacity}\label{sec:managerial}
	
	In this section, we assess the impact of two attributes in the room capacity: i) installing plexiglass divisions; and ii) considering the sitting sense of the customers to relax the distancing constraint when the sitting senses of two customers form an angle of 180$^{\circ}$. For the sake of conciseness, we limit our analysis to the problems of class $15 \times 15$, which are the largest problems that our model can consistently solve with little computational effort for all model variants.
	
\begin{figure}[!ht]
		\centering
		\includegraphics[trim=1.8cm 0.5cm 1cm 2cm, clip=true, scale=0.5]{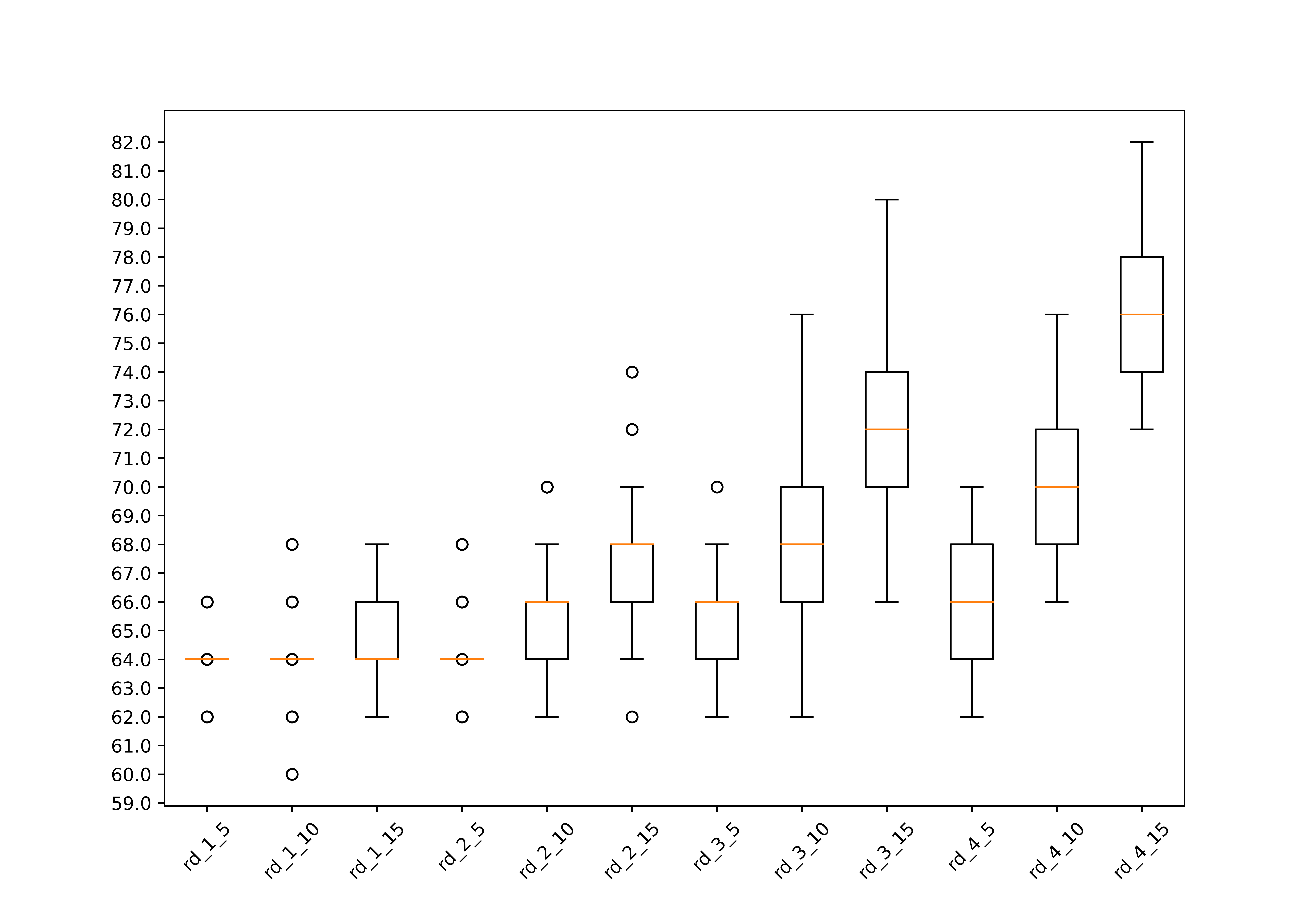}
		\caption{Capacity after installing plexiglass divisions on problem of size $15\times 15$}
		\label{fig:rd_walls_no_back_back_allowed}
	\end{figure}

	In our first experiment, we analyze the impact of installing plexiglass divisions on a square area, for different number and sizes of the plexiglass walls. In the case of the problem
	\texttt{15\_15\_nowalls}, the \texttt{Baseline} setting provides an optimal solution of 64 people sitting in the room.
	In the box plot of Figure \ref{fig:rd_walls_no_back_back_allowed} we report, for each $t\in\{1, 2, 3, 4\}$ and for every $w\in\{5, 10, 15\}$, the median value of the capacity, lower and upper quartiles, the minimum and maximum values, and the outliers when considering our model with setting \texttt{Plexiglass} for the same problem. Let us recall that 100 different random placements are considered for each setting. 
	
\begin{figure}[!ht]
		\centering
		\begin{subfigure}{0.5\textwidth}
			\centering
			\includegraphics[scale=0.07]{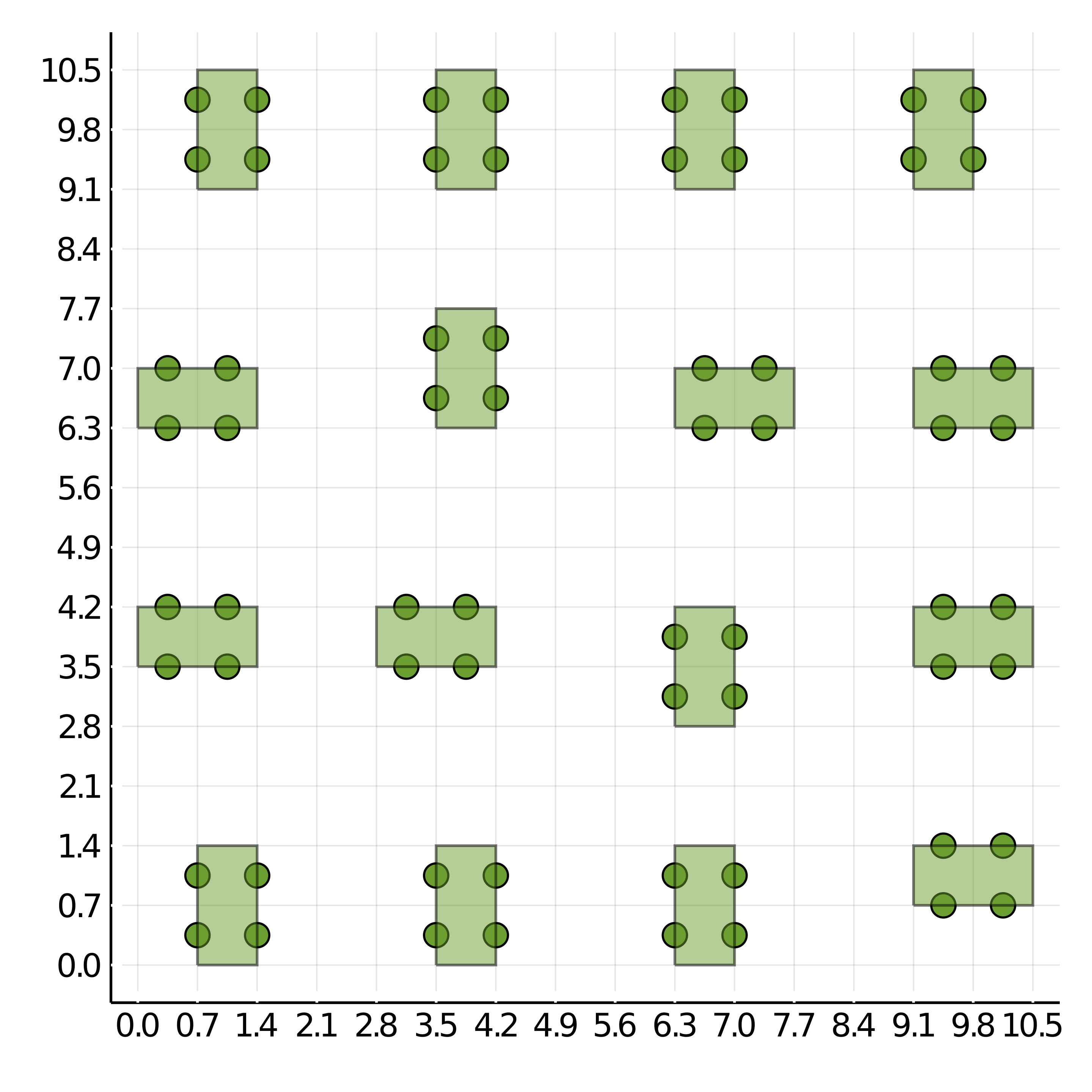}
			\caption{\texttt{Baseline} -- Instance \texttt{15\_15}}
			\label{fig:no_back_back}
		\end{subfigure}%
		\begin{subfigure}{0.5\textwidth}
			\centering
			\includegraphics[scale=0.07]{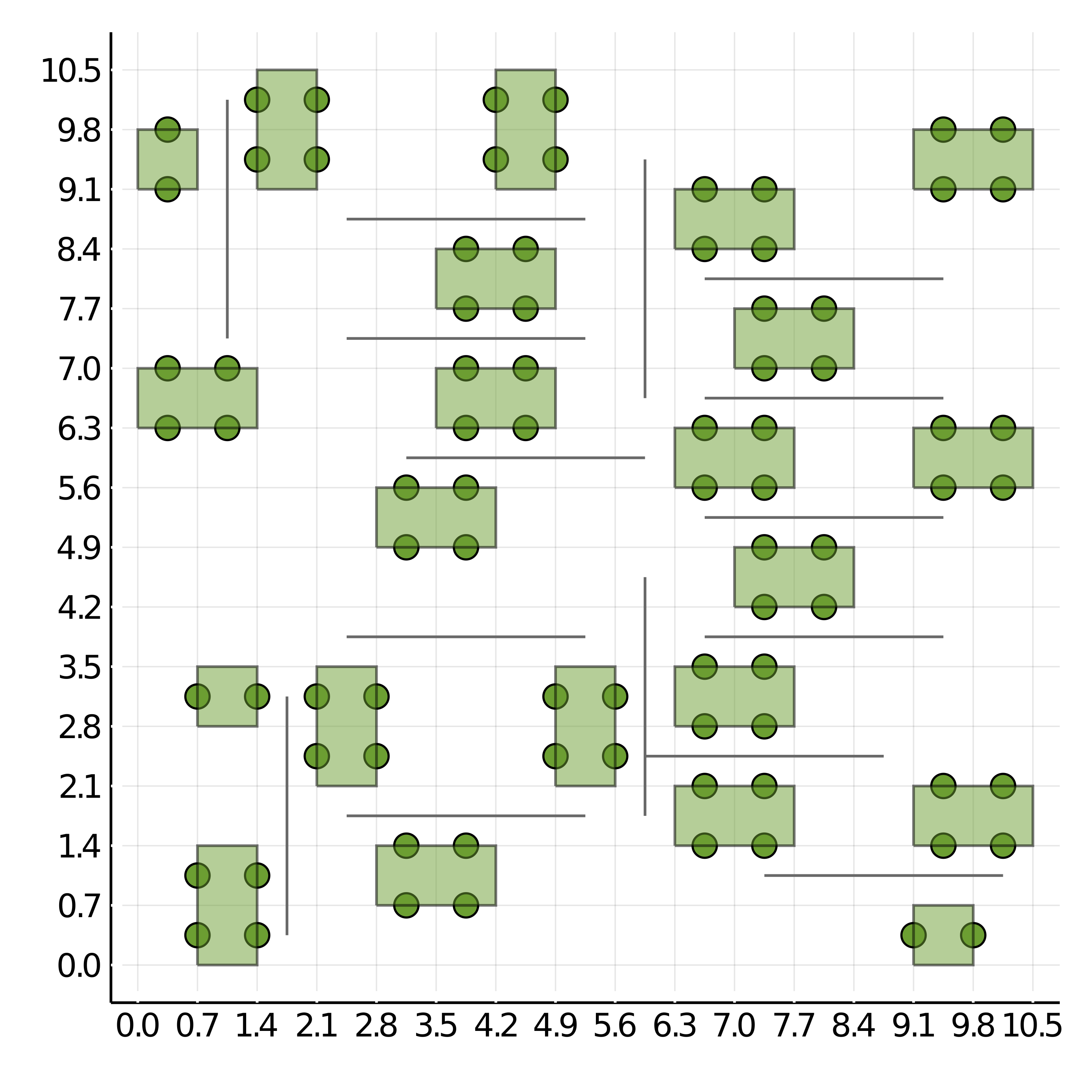}
			\caption{Best layout by \texttt{Plexiglass} -- Instance \texttt{15\_15}}
			\label{fig:max_capacity_no_back_back}
		\end{subfigure}%
		\caption{Different layouts produced by the \texttt{Baseline} vs \texttt{Plexiglass} settings
			\label{fig:baseline_vs_plexi_example}}
	\end{figure}

	We observe that the plexiglass divisions are effective at increasing the capacity, as the median values range often between 64 and 76. However, when one looks at the bottom outliers, we also observe that a careless placement of the divisions may result in no increase of the capacity at all. But also this plot suggests a positive result, which is that a careful placement of the divisions may result in a substantial increase of the room capacity (in this case, we observe an increase in the capacity of up to a 28\% with respect to the \texttt{Baseline} setting). We remark that in these problems, the walls have been placed at random, which suggests that an even more careful placement of the divisions may result in an even higher occupancy. This is a matter that definitely deserves further investigation. 
	In Figure \ref{fig:baseline_vs_plexi_example}, we plot the two optimal solutions provided by the \texttt{Baseline} and \texttt{Plexiglass} settings for this problem, the former capable of sitting 64 persons, while the latter attaining a maximum occupancy of 82 guests.
	
\begin{figure}[!ht]
		\centering
		\includegraphics[scale=0.07]{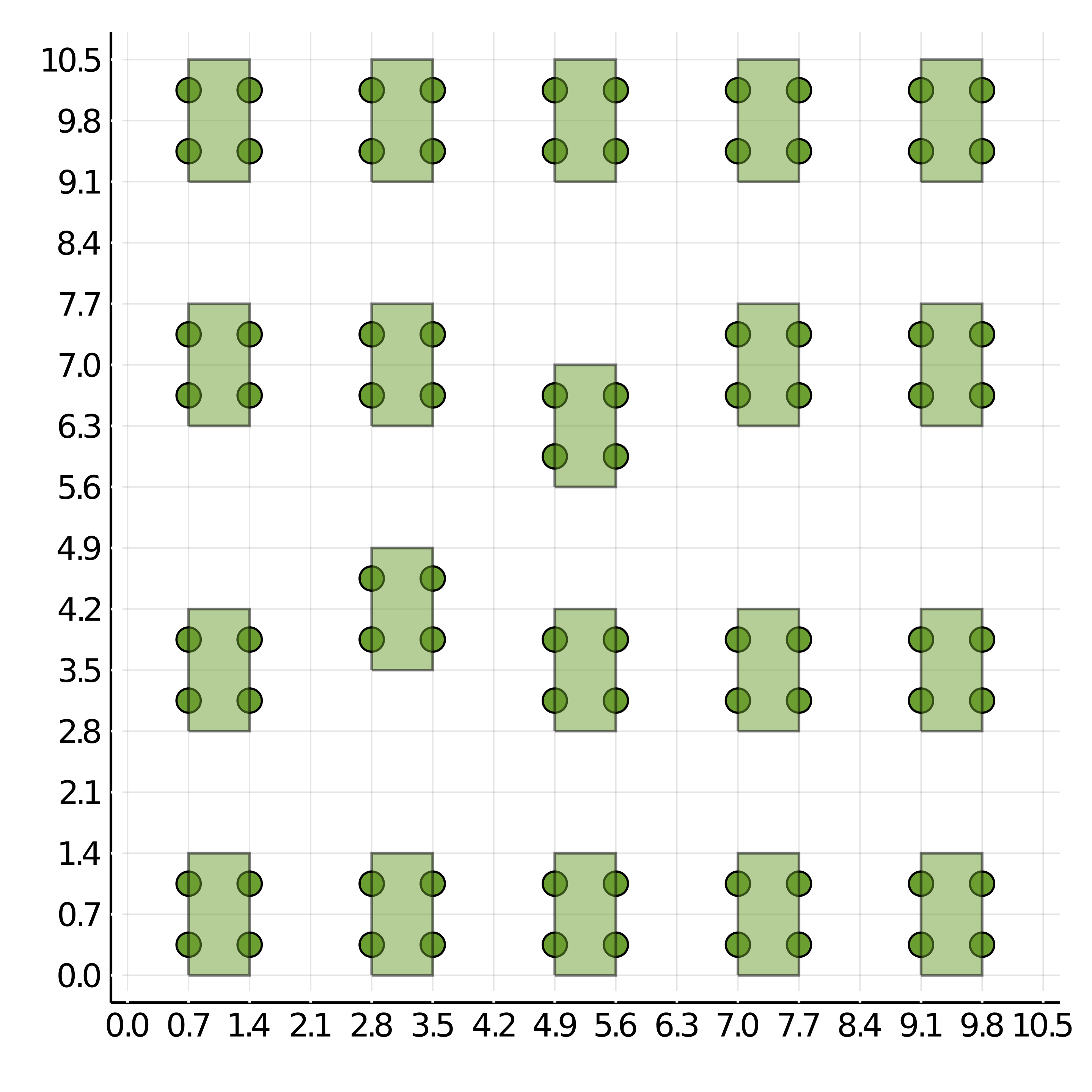}
		\caption{\texttt{SittingSense} -- Instance \texttt{15\_15}}
		\label{fig:layout_sitting_sense}	
	\end{figure}

	In our second experiment we analyze the impact of considering the sitting sense of the customers in the room capacity. For this, we consider settings \texttt{Baseline} and \texttt{SittingSense} on the same problem of size $15\times 15$. In Figure \ref{fig:layout_sitting_sense} we illustrate the optimal solution attained for the setting \texttt{SittingSense}, which results in an optimal occupancy of 80 guests, comparable to the effect of installing the plexiglass divisions. This is more than a 25\% 	larger than the maximum occupancy attained in the \texttt{Baseline} regime.
	
\begin{figure}[!ht]
		\centering
		\includegraphics[scale=0.5]{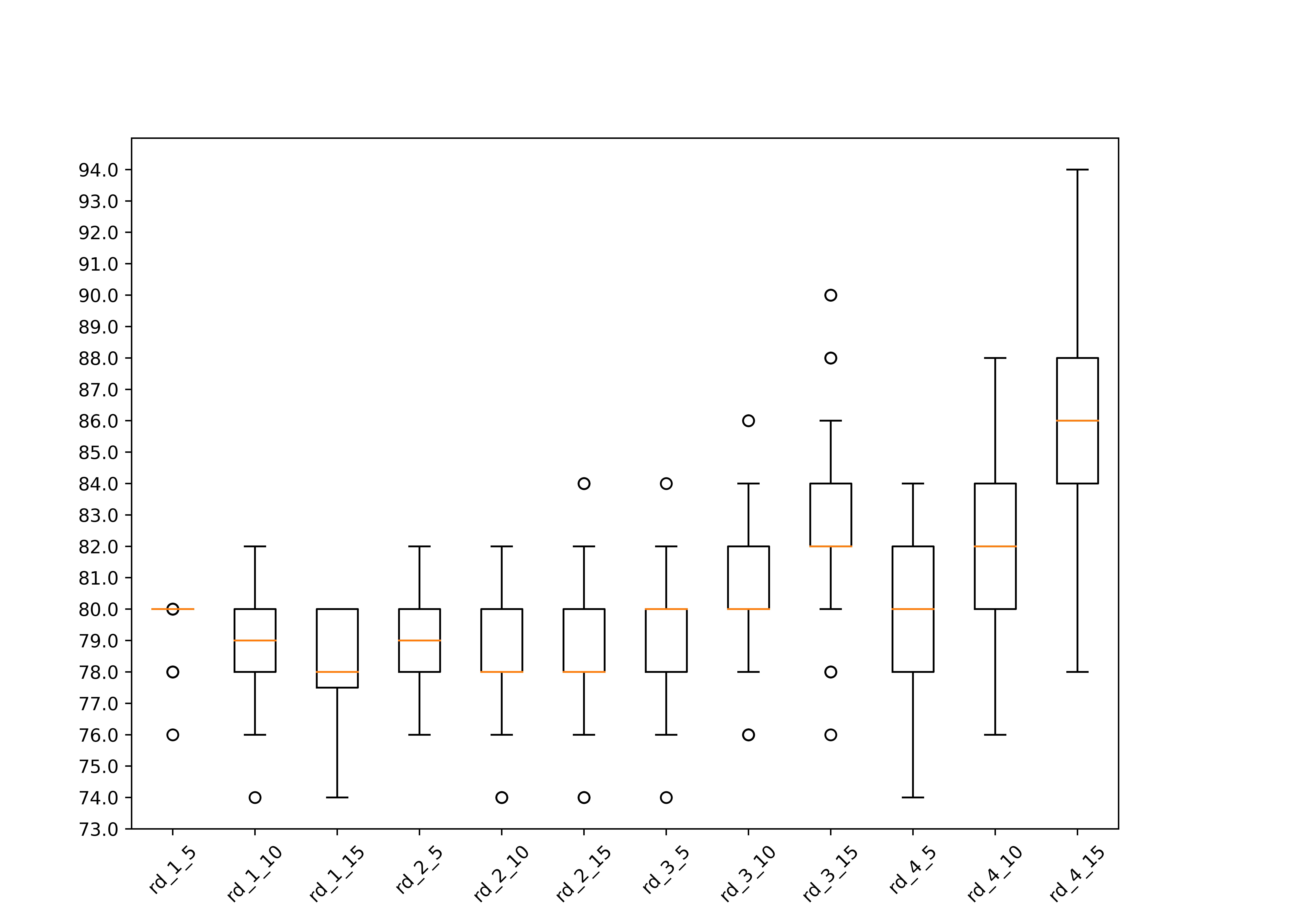}
		\caption{\texttt{Plexiglass+SittingSense}}
		\label{fig:plexi+stting_boxplot}
	\end{figure}

\begin{figure}[!ht]
		\centering
		\includegraphics[scale=0.07]{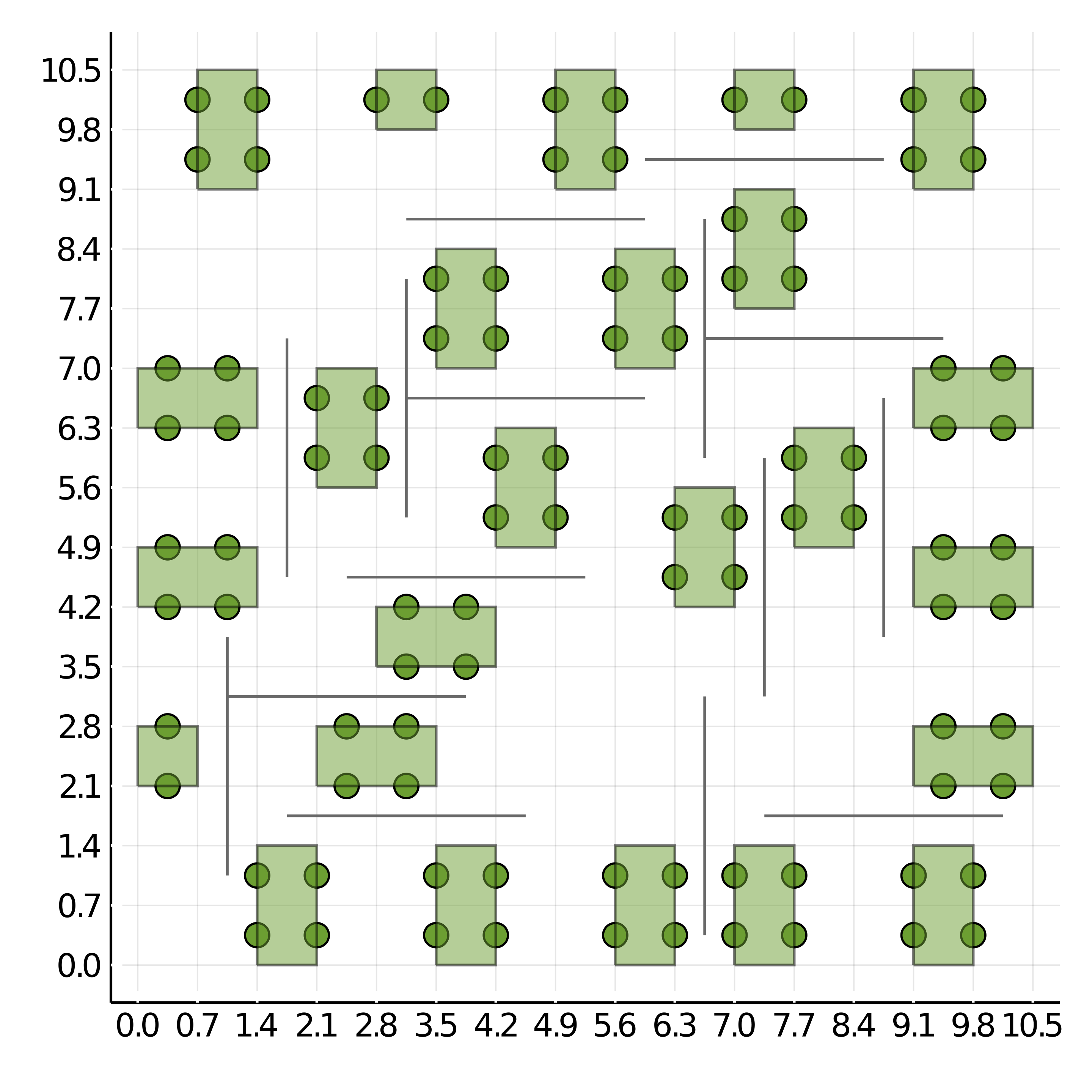}
		\caption{Best layout found by \texttt{Plexiglas+SittingSense} -- Instance \texttt{15\_15}}
		\label{fig:max_capacity_plexi+sitting}	
	\end{figure}

	Our third experiment aims at combining the installation of plexiglass divisions and the consideration of the sitting sense of customers in the distancing requirements to assess if the maximum occupancy can be increased further. We consider then the setting \texttt{Plexiglas+SittingSense} and compare it against the \texttt{Plexiglass} and the \texttt{SittingSense} variants. Figure \ref{fig:plexi+stting_boxplot} reports a box plot of the occupancy attained for the various random placements of the divisions.
	The result of combining these two features make the occupancy less sensitive to the location of the plexiglass divisions, and produces an increase in the medians that now range between 78 and 86. The maximum, however, is attained for the largest number and size of walls, of value 94. This is more than 14.6\% and 17\% superior to the largest occupancy achieved, respectively, by the \texttt{Plexiglass} and \texttt{SittingSense} settings separately. This shows that it is indeed relevant to consider these two features simultaneously when designing the layout of restaurant dining rooms. In Figure \ref{fig:max_capacity_plexi+sitting} we plot the best possible layout (of value 94) found by the \texttt{Plexiglass+SittingSense} variant of the problem.
	
	\label{fig:max_capacities}
	
	Our last experiment in this section aims at motivating further studies in the problem of locating plexiglass divisions. We consider the problem resulting from locating five plexiglasss divisions on three rows each, for a total of 15 plexiglass divisions, on a grid of size $15\times 15$, using the uniform distribution of walls as described in Section \ref{sec:instances}.	One would expect that, because of being more uniformly distributed, they would entail better occupancies than the problems in which divisions are randomly placed.
	The truth is, this is only true on average, but the best possible random placement results in a better occupancy than the one obtained from the uniform placement.
	The	random placement attains maximums of 82 and 94	respectively for the settings \texttt{Plexiglass} and \texttt{Plexiglass+SittingSense}, while the uniform placement	attains occupancies of 84 and 92 guests for the same settings.
	This does not mean that the random placement is always better, as these comparisons are only valid for the best solution found among all possible random placements, but suggests that the subject of locating plexiglass divisions in the best optimal way is not as simple as following an uniform placement as the one described in Section \ref{sec:instances}.
	
	\subsection{Scalability of the model}\label{sec:sensitivity}
	
	In this section, we analyze the scalability of the proposed model to three features, namely: 1) the size of the grid; 2) the presence or absence of plexiglass divisions; and 3) the consideration of (or lack thereof) the sitting sense of customers in the scalability of our model.
	
	Tables \ref{tab:no_walls_back_back_allowed} and \ref{tab:no_walls_back_back_not_allowed} show the performance of the model when solving problems representing rooms with no plexiglass divisions. We compare the impact of taking into account the sitting sense of the customers. For this, we consider the model settings \texttt{Baseline} and \texttt{SittingSense}. For each instance, we impose a time limit of 24h. The results are presented in Tables \ref{tab:no_walls_back_back_allowed} and \ref{tab:no_walls_back_back_not_allowed}. For each problem, we provide the number of rows and columns in the grid (\textbf{\# rows grid} and \textbf{\# cols grid}), the number of persons sitting in the room (\textbf{\# persons}), the computing time required to solve the model (\textbf{T(s)}), and the number of variables and constraints (\textbf{\# vars} and \textbf{\# edges}) in the corresponding model. The information of these two last columns gives us an idea about the size of the conflict graph associated with the problem. Finally, for the unsolved instances, due to time limit, we report the corresponding optimality gap (\textbf{gap$_{opt}$}) at the moment when the solution process stops. Optimality gaps are computed as $(\mathbf{{\bar{Z}}}-\mathbf{{\underline{Z}}})/\mathbf{{\underline{Z}}}$, where $\mathbf{{\underline{Z}}}$ and $\mathbf{{\bar{Z}}}$ represents primal and dual bounds, respectively.
	
	Regarding the results, as one might expect, the setting \texttt{SittingSense} allows more people to be placed in each room. Moreover, allowing people to sit closer to each other yields a smaller conflict graph, which is due to the smaller number of incompatibilities. This aspect impacts the computing times required to solve each problem. The model that is solved over the conflict graph associated with \texttt{SittingSense} was capable of solving all the instances associated with grids with sizes going up to $30\times30$ in less than 50 seconds.
	In turn, the model arising from \texttt{Baseline} was not capable of solving instances with more than 15 rows and 15 columns within the time limit of 24h imposed. In addition, we also observe that the grid size substantially affects the computing times for the \texttt{Baseline} model as we perceive an increase of orders of magnitude in the computing times when passing from a grid of size $15\times 15$ to one of size $20\times 20$. This sensitivity is substantially reduced when one considers the sitting sense of the customers in the modeling. The formulation starts to be too time-consuming when tackling instances with grids with size superior to $30\times 30$. Yet, the optimality gap attained when solution process stops is small.
	
	\begin{table}[htb]
			\small
		\centering
		\caption{Results obtained by considering \texttt{SittingSense}}
		\begin{tabular}{cccccccc}
			\hline
			\textbf{Instance name} & \textbf{\# rows grid} & \textbf{\# cols grid} & \textbf{\# vars} & \textbf{\# edges} & \textbf{\# persons} & \textbf{gap$_{opt.}$} & \textbf{T(s)} \\ 
			\hline
			\texttt{5\_5\_nowalls}  & 5 & 5 & 54 & 1,313 & 6 &  --  & 0.2 \\ 
			\texttt{10\_10\_nowalls}  & 10 & 10 & 304 & 15,118 & 36 &  --   & 0.3 \\ 
			\texttt{15\_15\_nowalls}  & 15 & 15 & 754 & 43,673 & 80 &  --   & 1.3 \\ 
			\texttt{20\_20\_nowalls}  & 20 & 20 & 1404 & 86,978 & 132 &  --   & 24.5 \\ 
			\texttt{25\_25\_nowalls}  & 25 & 25 & 2254 & 149,033 & 214 &  --   & 49.2 \\ 
			\texttt{30\_30\_nowalls}  & 30 & 30 & 3304 & 217,838 & 320 &  --   & 13 \\ 
			\texttt{35\_35\_nowalls}$^*$  & 35 & 35 & 4554 & 305,393 & 414$^{**}$ &  1.60\%  & 86,400 \\ 
			\texttt{40\_40\_nowalls}$^*$  & 40 & 40 & 6004 & 407,698 & 532$^{**}$ &  4.18\%  & 86,400 \\ 
			\hline
			\multicolumn{8}{l}{\scriptsize $^{*}$ The instance could not be solved within the time limit of 24h.} \\
			\multicolumn{8}{l}{\scriptsize $^{**}$ Best primal bound obtained after 24h of computing time.} \\
			
		\end{tabular}
		\label{tab:no_walls_back_back_allowed}
	\end{table}
	
	\begin{table}[htb]
			\small
		\centering
		\caption{Results obtained by considering \texttt{Baseline}}
		\begin{tabular}{cccccccc}
			\hline
			\textbf{Instance name} & \textbf{\# rows grid} & \textbf{\# cols grid} & \textbf{\# vars} & \textbf{\# edges} & \textbf{\# persons} & \textbf{gap$_{opt.}$} & \textbf{T(s)} \\ 
			\hline
			\texttt{5\_5\_nowalls}  & 5 & 5 & 54 & 1313 & 6 &  --  & 0.2 \\ 
			\texttt{10\_10\_nowalls}  & 10 & 10 & 304 & 17028 & 28 &  --  & 0.4 \\ 
			\texttt{15\_15\_nowalls}  & 15 & 15 & 754 & 49893 & 64 &  --  & 1.5 \\ 
			\texttt{20\_20\_nowalls}$^*$  & 20 & 20 & 1,404 & 99,908 &  102$^{**}$  &  7.38\%  & 86400 \\ 
			\texttt{25\_25\_nowalls}$^*$  & 25 & 25 & 2,254 & 167,073 &  162$^{**}$  &  8.72\%  & 86400 \\ 
			\texttt{30\_30\_nowalls}$^*$  & 30 & 30 & 3,304 & 251,388 &  240$^{**}$  &  6.58\%  & 86400 \\ 
			\texttt{35\_35\_nowalls}$^*$  & 35 & 35 & 4,554 & 352,853 &  324$^{**}$  &  8.24\% & 86400 \\ 
			\texttt{40\_40\_nowalls}$^*$  & 40 & 40 & 6,044 & 471,468 &  400$^{**}$  &  14.74\% & 86400 \\ 
			\hline
			\multicolumn{8}{l}{\scriptsize $^{*}$ The instance could not be solved within the time limit of 24h.} \\
			\multicolumn{8}{l}{\scriptsize $^{**}$ Best primal bound obtained after 24h of computing time.} \\
		\end{tabular}
		\label{tab:no_walls_back_back_not_allowed}
	\end{table}
	
	Due to its structure, formulation \eqref{eq:objfunction}--\eqref{eq:variables} may present a lot of symmetry. As a consequence, it may struggle to close gaps when solving certain instances. Hence, in order to allow us to have some insights on the performance of the formulation with respect to its capability of finding feasible solutions and to proving optimality, we report optimality gaps obtained after 1h for the unsolved instances in Tables \ref{tab:no_walls_back_back_allowed} and \ref{tab:no_walls_back_back_not_allowed}.
	In Table \ref{tab:optimality_gaps}, for the \texttt{Baseline} setting, we observe that, except for instance \texttt{40\_40\_nowalls}, the formulation is capable of finding, after 1h, feasible solutions with 10\% from optimality. For the \texttt{Plexiglass+SittingSense} setting, in turn, feasible solutions with 5\% from optimality were found.
	
	Notice that for instances the \texttt{15\_15\_nowalls} and \texttt{20\_20\_nowalls}, in the \texttt{Baseline} setting, and the instance \texttt{35\_35\_nowalls}, in the \texttt{Plexiglass+SittingSense} setting, the obtained primal bounds after 1h did not change until the method stopped after 24h of execution (Tables \ref{tab:no_walls_back_back_allowed} and \ref{tab:no_walls_back_back_not_allowed}).
	
	\begin{table}[!htbp]
		\caption{Performance of the formulation on its ability to close gaps}
		\centering
		\begin{tabular}{ccccccc}
			\hline
			\textbf{Setting} & \textbf{Instance name} & \textbf{\# vars} & \textbf{\# edges} & $\mathbf{{\underline{Z}}}$ & $\mathbf{{\bar{Z}}}$ & \textbf{gap$_{opt}$} \\ 
			\hline
			\multirow{5}{*}{\texttt{Baseline}} & \texttt{20\_20\_nowalls} & 1,404 & 99,908 & 102 & 112.52 & 10.31 \% \\ 
			& \texttt{25\_25\_nowalls} & 2,254 & 167,073 & 162 & 178.26 & 10.04 \% \\
			& \texttt{30\_30\_nowalls} & 3,304 & 251,388 & 236 & 257.95 & 9.39 \% \\
			& \texttt{35\_35\_nowalls} & 4,554 & 352,853 & 324 & 351.95 & 8.63 \% \\
			& \texttt{40\_40\_nowalls} & 6,004 & 471,468 & 398 & 459.65 & 15.49 \% \\
			\hline
			\multirow{2}{*}{\texttt{Plexiglass+SittingSense}}  & \texttt{35\_35\_nowalls} & 4,554 & 305,393 & 414 & 423.66 & 2.23 \%\\
			& \texttt{40\_40\_nowalls} & 6,004 & 407,698 & 530 & 556.63 & 5.03 \%\\ \hline
		\end{tabular}
		\label{tab:optimality_gaps}
	\end{table}

\begin{figure}
		\includegraphics[scale=0.6]{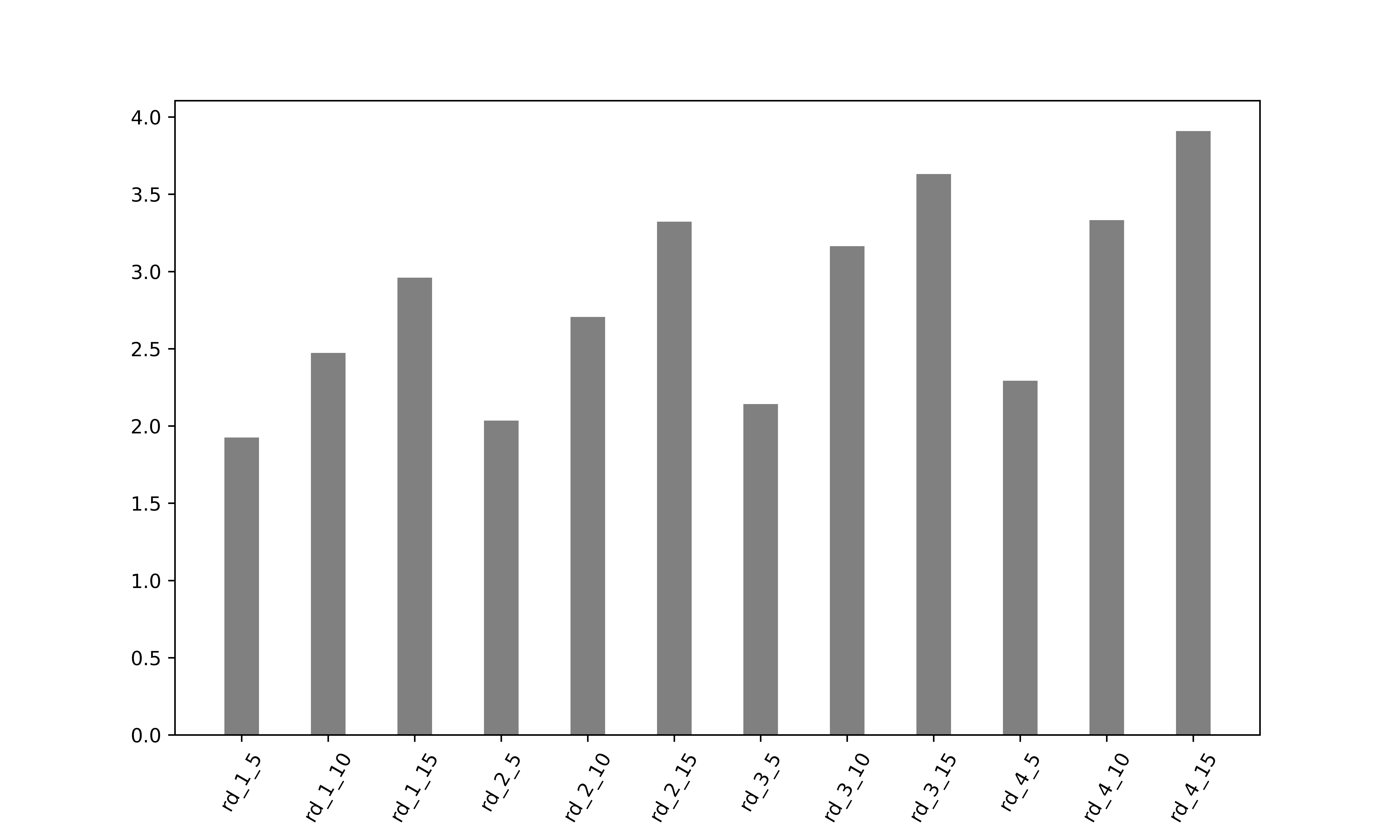}
		\caption{Average computing times for \texttt{Plexiglass}\label{fig:barplot:plexi}}
	\end{figure}

	Finally, we report computing times obtained when solving instances considering plexiglass walls.
	The bar plot depicted in Figure \ref{fig:barplot:plexi} reports the computing times (in seconds) required by our algorithm to solve the model associated with the problem setting \texttt{Plexiglass}. Please recall that the computing time taken by the \texttt{Baseline} setting for this instance was of 1.5 seconds. In this bar plot, we observe that the average times of all different configurations require between two and four seconds.
	The computing times are lower, but of the same order of magnitude than those taken by the \texttt{Baseline} model.
	
		\subsection{Comparison against constructive heuristics}\label{sec:heuristics}
		
		In this section, we propose two constructive heuristics and we compare their performance to the exact method that solves formulation  \eqref{eq:objfunction}--\eqref{eq:variables}.
		Much like the exact method, both heuristics rely on the enumeration of sitting configurations.
		The first heuristic (\texttt{close\_corner}) consists in placing tables as close as possible to one of the corners of the room.
		More specifically, in our implementation, the reference corner was the origin of the axis $x$ and $y$.
		We sort the sitting configurations in non-decreasing order according to the distance of the centers of the corresponding tables to the chosen corner.
		Then, at each iteration, we select and add to the solution the configuration for which the corresponding table is closest to the chosen corner, provided that it is compatible with all the other sitting configurations already accommodated.
		In the case where two sitting configurations are located at the same distance from the corner, the tie-break is done by favoring the sitting configuration accommodating more customers.
		This approach enables to implicitly place, at a time, one table accommodating as many customers as possible, and that is as close as possible to the ones already placed.
		The second heuristic (\texttt{random}) randomly selects sitting configurations and tries to place them (add to the solution) if they are not incompatible with the ones already placed. For \texttt{random}, the reported results are obtained by computing an average of five executions of the method.
		
		Initially, we compare the performance of the heuristics when dealing with instances corresponding to rooms with plexiglass walls. In Table \ref{tab:heuristics_instances_with_walls}, we report the average gaps ({\% gap}) of the solutions obtained by each heuristic with respect the exact method (\texttt{exact}), as well as the average computing times required by each heuristic (\textbf{T(s)}), when considering settings \texttt{Baseline} and \texttt{Plexiglass+SittingSense}.
	
	\begin{table}[htbp]
		\caption{Performance of the proposed method against constructive heuristics}
		\centering
		\footnotesize
		\renewcommand{\tabcolsep}{0.2cm}	
		\begin{tabular}{cccccccccccc}
			\hline
			\multirow{3}{*}{\textbf{Instance}} & \multicolumn{5}{c}{\texttt{Baseline}} && \multicolumn{5}{c}{\texttt{Plexiglass+SittingSense}}\\
			\cline{2-6}
			\cline{8-12}
			& \multicolumn{2}{c}{\texttt{exact} $\times$ \texttt{close\_corner}} && \multicolumn{2}{c}{\texttt{exact} $\times$ \texttt{random}} && \multicolumn{2}{c}{\texttt{exact} $\times$ \texttt{close\_corner}} && \multicolumn{2}{c}{\texttt{exact} $\times$ \texttt{random}} \\		 
			\cline{2-3}
			\cline{5-6}
			\cline{8-9}
			\cline{11-12}
			& \textbf{\% gap} & \textbf{T(s)} & & \textbf{\% gap} & \textbf{T(s)} && \textbf{\% gap} & \textbf{T(s)} & & \textbf{\% gap} & \textbf{T(s)} \\
			\hline
			\texttt{15\_15\_1\_5} & 3.47 & 1.9 &  & 41.53 & 1.9 &  & 3.18 & 2.0 &  & 45.56 & 2.0 \\ 
			\texttt{15\_15\_1\_10} & 5.77 & 2.4 &  & 41.87 & 2.4 &  & 5.85 & 2.3 &  & 45.07 & 2.3 \\ 
			\texttt{15\_15\_1\_15} & 7.83 & 2.7 &  & 40.95 & 2.7 &  & 7.94 & 2.6 &  & 43.84 & 2.6 \\ 
			\hline
			\texttt{15\_15\_2\_5} & 5.91 & 2.0 &  & 40.39 & 1.9 &  & 6.69 & 1.9 &  & 43.69 & 1.9 \\ 
			\texttt{15\_15\_2\_10} & 9.43 & 2.6 &  & 40.08 & 2.6 &  & 10.04 & 2.4 &  & 42.52 & 2.4 \\ 
			\texttt{15\_15\_2\_15} & 11.20 & 3.1 &  & 39.87 & 3.1 &  & 12.34 & 2.9 &  & 41.39 & 2.8 \\ 
			\hline
			\texttt{15\_15\_3\_5} & 7.43 & 2.0 &  & 40.24 & 2.0 &  & 7.34 & 2.0 &  & 43.01 & 2.0 \\ 
			\texttt{15\_15\_3\_10} & 9.88 & 2.9 &  & 39.84 & 2.8 &  & 10.94 & 2.6 &  & 41.86 & 2.7 \\ 
			\texttt{15\_15\_3\_15} & 11.62 & 3.5 &  & 38.71 & 3.4 &  & 12.60 & 3.0 &  & 40.29 & 3.1 \\ 
			\hline
			\texttt{15\_15\_4\_5} & 7.30 & 2.1 &  & 40.81 & 2.1 &  & 7.27 & 2.0 &  & 43.44 & 2.1 \\ 
			\texttt{15\_15\_4\_10} & 8.86 & 3.0 &  & 37.88 & 3.0 &  & 11.25 & 2.8 &  & 39.40 & 2.8 \\ 
			\texttt{15\_15\_4\_15} & 8.39 & 3.6 &  & 36.62 & 3.6 &  & 10.75 & 3.2 &  & 37.34 & 3.2 \\ 
			\hline
		\end{tabular}
		\label{tab:heuristics_instances_with_walls}
	\end{table}
	
	In Table \ref{tab:heuristics_instances_with_walls}, we observe that for the instances associated with grids of size $15\times15$, the average computing times of \texttt{exact} (Figure \ref{fig:barplot:plexi}) are very similar to those required by the heuristic methods. Regarding the quality of the solutions, for the \texttt{Baseline} setting, the reported gaps vary from 3.47\% to 11.62\%, when comparing \texttt{exact} and \texttt{close\_corner}, and between 36.62\% and 41.87\% when comparing \texttt{exact} and \texttt{random}. We observe that the performance of \texttt{close\_corner} is reasonable, especially when tackling problems with a low number of short plexiglass walls. In other words, the performance of \texttt{close\_corner} deteriorates as the number of plexiglass walls and their lengths increase. The \texttt{random} method, in turn, did not perform well. Similar conclusions could be drawn by analyzing the results obtained for \texttt{Plexiglass+SittingSense}.
		
	Now, we analyze the performance of the constructive heuristics when tackling problems representing rooms with no plexiglass divisions. In Table \ref{tab:heuristic_nowalls} we report the average computing time required by the heuristics to solve each problem (\textbf{T(s)}), as well as the value corresponding to the solution obtained (\textbf{value}). Moreover, in order to have a reference for comparison, we report solution values obtained by the \texttt{exact} method. When the optimal solution is not available, we report the best primal solution obtained after 1h of execution time. From the results in Table \ref{tab:heuristic_nowalls}, we observe that \texttt{close\_corner} performs very well for both \texttt{Baseline} and \texttt{SittingSense} settings, whereas \texttt{random} performs poorly. This behavior is consistent with the one observed for the instances corresponding to rooms with walls (Table \ref{tab:heuristics_instances_with_walls}). For the \texttt{Baseline} setting, \texttt{close\_corner} was capable of finding, in less than 10s, solutions that are very close to those obtained by the \texttt{exact} approach in the time limit of 1h. In fact, sometimes \texttt{close\_corner} was able to provide optimal solutions for some problems \texttt{5\_5\_nowalls}, \texttt{10\_10\_nowalls}, and \texttt{15\_15\_nowalls}, or even to find solutions that are better than the ones obtained by \texttt{exact} after 1h, and similar to the ones obtained after 24h of computing (\texttt{40\_40\_nowalls}). For \texttt{SittingSense}, \texttt{close\_corner} was capable of finding the optimal solution for the instance \texttt{30\_30\_nowalls}. This good performance of \texttt{close\_corner} when tackling instances without walls was expected, since the method shows the best performance in Table \ref{tab:heuristics_instances_with_walls} when solving problems with just a few and short plexiglass walls.
	
	\begin{table}[htbp]
		\caption{Heuristic solutions }
		\small
		\centering
		\begin{tabular}{cccccccccccc}
			\hline
			\multirow{3}{*}{\textbf{Instance}}& \multicolumn{5}{c}{\texttt{Baseline}}  &  & \multicolumn{5}{c}{\texttt{SittingSense}}  \\ 
			\cline{2-6}
			\cline{8-12}
			& \multicolumn{2}{c}{\texttt{close\_corner}} & \multicolumn{2}{c}{\texttt{random}}  & \texttt{exact} & & \multicolumn{2}{c}{\texttt{close\_corner}} & \multicolumn{2}{c}{\texttt{random}} & \texttt{exact}  \\ 
			\cline{2-6}
			\cline{8-12}
			& \textbf{value} & \textbf{T(s)} & \textbf{value} & \textbf{T(s)} & \textbf{value} & & \textbf{value} & \textbf{T(s)} & \textbf{value} & \textbf{T(s)} & \textbf{value}\\ 
			\hline
			\texttt{5\_5\_nowalls} & 6 & 0.0 & 4 & 0.0 & 6 &  & 6 & 0.0 & 4 & 0.0 & 6 \\ 
			\texttt{10\_10\_nowalls} & 28 & 0.2 & 16 & 0.2 & 28 &  & 36 & 0.2 & 21 & 0.2 & 36 \\ 
			\texttt{15\_15\_nowalls} & 64 & 0.7 & 34 & 0.6 & 64 &  & 80 & 0.8 & 41 & 0.9 & 80 \\ 
			\texttt{20\_20\_nowalls} & 100 & 1.4 & 63 & 1.3 & 102$^*$ &  & 120 & 1.6 & 74 & 1.8 & 132 \\ 
			\texttt{25\_25\_nowalls} & 156 & 2.1 & 89 & 2.2 & 162$^*$ &  & 208 & 2.7 & 106 & 2.8 & 214 \\ 
			\texttt{30\_30\_nowalls} & 238 & 3.8 & 135 & 3.4 & 236$^*$ &  & 320 & 4.4 & 169 & 4.3 & 320 \\ 
			\texttt{35\_35\_nowalls} & 324 & 4.8 & 183 & 4.9 & 324$^*$ &  & 396 & 6.2 & 223 & 6.0 & 414$^*$ \\ 
			\texttt{40\_40\_nowalls} & 400 & 6.5 & 233 & 6.4 & 398$^*$ &  & 520 & 8.4 & 280 & 8.1 & 530$^*$ \\ 
			\hline
			\multicolumn{12}{l}{\scriptsize $^*$ Best primal bounds obtained after 1h.} \\
		\end{tabular}
		\label{tab:heuristic_nowalls}
	\end{table}

	\section{Conclusions}\label{sec:conclusions}
	
	We have introduced a mathematical model and two constructive heuristics for the problem of designing the optimal layout of a restaurant dining room under several attributes relevant in practice at the times of the current pandemic, such as the arbitrary topology of the dining room, an arbitrary configuration of the tables, the sitting sense of customers, and the presence of (or lack thereof) plexiglass divisions.
	
	The model introduced is extremely simple, as it relies on some basic concepts from the mathematical optimization literature. Therefore, it can be easily implemented and adapted to other settings. Our model was capable of handling quite large problems, on grids of up to $40\times 40$ squares, within low computing times provided that the sitting sense of the customers was not ignored.
	
	We have also performed two types of analysis, one aiming at assessing the impact of some of the attributes in the room capacity, and another aiming at assessing the computational behavior of our model.
	Our findings suggest that the installation of plexiglass divisions and the consideration of the sitting sense of customers explicitly in the modeling can entail increases of
	almost 30\%
	of the capacity of the dining room. Moreover, both attributes, when considered simultaneously, can help at increasing the room capacity even further. Our analysis also allowed us to conclude that an uniform placement of the plexiglass divisions is not necessarily optimal, leaving room for a future study on the optimal placement of plexiglass divisions.

	From the computational standpoint, we have showed that our model scales well when plexiglass divisions and when the sitting sense of the customers are explicitly considered, as opposed to a baseline variant of our model in which those attributes are not present, and where our model's scalability is much worse. One of our constructive heuristics, on the other hand, finds good quality solutions within a fraction of the computing times for problems with few obstacles or plexiglass divisions.
	
\section*{Acknowledgments}

We thank the anonymous reviewer and the Associate Editor whose comments and suggestions helped us improve the quality of our manuscript. C. Contardo thanks the Natural Sciences and Engineering Research Council of Canada (NSERC) for its financial support through Grant no. 2020-06311.

\bibliographystyle{elsarticle-harv}
\bibliography{library}

\begin{thebibliography}{25}
\expandafter\ifx\csname natexlab\endcsname\relax\def\natexlab#1{#1}\fi
\providecommand{\url}[1]{\texttt{#1}}
\providecommand{\href}[2]{#2}
\providecommand{\path}[1]{#1}
\providecommand{\DOIprefix}{doi:}
\providecommand{\ArXivprefix}{arXiv:}
\providecommand{\URLprefix}{URL: }
\providecommand{\Pubmedprefix}{pmid:}
\providecommand{\doi}[1]{\href{http://dx.doi.org/#1}{\path{#1}}}
\providecommand{\Pubmed}[1]{\href{pmid:#1}{\path{#1}}}
\providecommand{\bibinfo}[2]{#2}
\ifx\xfnm\relax \def\xfnm[#1]{\unskip,\space#1}\fi
\bibitem[{{Alvarez-Valdes} et~al.(2018){Alvarez-Valdes}, Carravilla and
  Oliveira}]{AlvarezValdes2018}
\bibinfo{author}{{Alvarez-Valdes}, R.}, \bibinfo{author}{Carravilla, M.},
  \bibinfo{author}{Oliveira, J.}, \bibinfo{year}{2018}.
\newblock \bibinfo{title}{Cutting and packing}, in: \bibinfo{editor}{Mart\'\i,
  R.}, \bibinfo{editor}{Pardalos, P.}, \bibinfo{editor}{Resende, M.} (Eds.),
  \bibinfo{booktitle}{Handbook of Heuristics}. \bibinfo{publisher}{Springer},
  pp. \bibinfo{pages}{931--977}.
\bibitem[{Araujo and Naimi(2020)}]{araujo2020spread}
\bibinfo{author}{Araujo, M.B.}, \bibinfo{author}{Naimi, B.},
  \bibinfo{year}{2020}.
\newblock \bibinfo{title}{Spread of {SARS-CoV-2} coronavirus likely to be
  constrained by climate}.
\newblock \bibinfo{journal}{medRxiv} .
\bibitem[{Bezanson et~al.(2017)Bezanson, Edelman, Karpinski and
  Shah}]{Bezanson2017}
\bibinfo{author}{Bezanson, J.}, \bibinfo{author}{Edelman, A.},
  \bibinfo{author}{Karpinski, S.}, \bibinfo{author}{Shah, V.B.},
  \bibinfo{year}{2017}.
\newblock \bibinfo{title}{Julia: A fresh approach to numerical computing}.
\newblock \bibinfo{journal}{SIAM Review} \bibinfo{volume}{59},
  \bibinfo{pages}{65--98}.
\bibitem[{Bezerra et~al.(2020)Bezerra, Leao, Oliveira and Santos}]{Bezerra2020}
\bibinfo{author}{Bezerra, V.M.R.}, \bibinfo{author}{Leao, A.A.S.},
  \bibinfo{author}{Oliveira, J.F.}, \bibinfo{author}{Santos, M.O.},
  \bibinfo{year}{2020}.
\newblock \bibinfo{title}{Models for the two-dimensional level strip packing
  problem -- {A} review and a computational evaluation}.
\newblock \bibinfo{journal}{Journal of the Operational Research Society}
  \bibinfo{volume}{71}, \bibinfo{pages}{606--627}.
\bibitem[{Côté and Iori(2018)}]{Cote2018}
\bibinfo{author}{Côté, J.F.}, \bibinfo{author}{Iori, M.},
  \bibinfo{year}{2018}.
\newblock \bibinfo{title}{The meet-in-the-middle principle for cutting and
  packing problems}.
\newblock \bibinfo{journal}{INFORMS Journal on Computing} \bibinfo{volume}{30},
  \bibinfo{pages}{646--661}.
\bibitem[{Dbouk and Drikakis(2020)}]{Dbouk2020}
\bibinfo{author}{Dbouk, T.}, \bibinfo{author}{Drikakis, D.},
  \bibinfo{year}{2020}.
\newblock \bibinfo{title}{On coughing and airborne droplet transmission to
  humans}.
\newblock \bibinfo{journal}{Physics of Fluids} \bibinfo{volume}{32},
  \bibinfo{pages}{053310}.
\bibitem[{Dube et~al.(2021)Dube, Nhamo and Chikodzi}]{Dube2020}
\bibinfo{author}{Dube, K.}, \bibinfo{author}{Nhamo, G.},
  \bibinfo{author}{Chikodzi, D.}, \bibinfo{year}{2021}.
\newblock \bibinfo{title}{{COVID-19 cripples global restaurant and hospitality
  industry}}.
\newblock \bibinfo{journal}{Current Issues in Tourism} \bibinfo{volume}{24},
  \bibinfo{pages}{1487--1490}.
\bibitem[{Dubejko and Stephenson(1995)}]{Dubejko199519}
\bibinfo{author}{Dubejko, T.}, \bibinfo{author}{Stephenson, K.},
  \bibinfo{year}{1995}.
\newblock \bibinfo{title}{Circle packing: Experiments in discrete analytic
  function theory}.
\newblock \bibinfo{journal}{Discrete \& Computational Geometry}
  \bibinfo{volume}{22}, \bibinfo{pages}{19--39}.
\bibitem[{Dunning et~al.(2017)Dunning, Huchette and Lubin}]{Dunning2017}
\bibinfo{author}{Dunning, I.}, \bibinfo{author}{Huchette, J.},
  \bibinfo{author}{Lubin, M.}, \bibinfo{year}{2017}.
\newblock \bibinfo{title}{{JuMP}: A modeling language for mathematical
  optimization}.
\newblock \bibinfo{journal}{SIAM Review} \bibinfo{volume}{59},
  \bibinfo{pages}{295--320}.
\bibitem[{Duque et~al.(2020)Duque, Morton, Singh, Du, Pasco and
  Meyers}]{Duque2020}
\bibinfo{author}{Duque, D.}, \bibinfo{author}{Morton, D.P.},
  \bibinfo{author}{Singh, B.}, \bibinfo{author}{Du, Z.},
  \bibinfo{author}{Pasco, R.}, \bibinfo{author}{Meyers, L.A.},
  \bibinfo{year}{2020}.
\newblock \bibinfo{title}{Timing social distancing to avert unmanageable
  {COVID-19} hospital surges}.
\newblock \bibinfo{journal}{Proceedings of the National Academy of Sciences of
  the United States of America} \bibinfo{volume}{117},
  \bibinfo{pages}{19873--19878}.
\bibitem[{Feng et~al.(2020)Feng, Marchal, Sperry and Yi}]{feng2020influence}
\bibinfo{author}{Feng, Y.}, \bibinfo{author}{Marchal, T.},
  \bibinfo{author}{Sperry, T.}, \bibinfo{author}{Yi, H.}, \bibinfo{year}{2020}.
\newblock \bibinfo{title}{Influence of wind and relative humidity on the social
  distancing effectiveness to prevent {COVID-19} airborne transmission: {A
  numerical study}}.
\newblock \bibinfo{journal}{Journal of Aerosol Science} \bibinfo{volume}{147},
  \bibinfo{pages}{105585}.
\bibitem[{Fischetti et~al.(2021)Fischetti, Fischetti and
  Stoustrup}]{Fischetti2020}
\bibinfo{author}{Fischetti, M.}, \bibinfo{author}{Fischetti, M.},
  \bibinfo{author}{Stoustrup, J.}, \bibinfo{year}{2021}.
\newblock \bibinfo{title}{Safe distancing in the time of {COVID-19}}.
\newblock \bibinfo{journal}{European Journal of Operational Research} .
\bibitem[{Freeman and Eykelbosh(2020)}]{freeman2020covid}
\bibinfo{author}{Freeman, S.}, \bibinfo{author}{Eykelbosh, A.},
  \bibinfo{year}{2020}.
\newblock \bibinfo{title}{{COVID-19} and outdoor safety: Considerations for use
  of outdoor recreational spaces}.
\newblock \bibinfo{journal}{National Collaborating Centre for Environmental
  Health} .
\bibitem[{Iori et~al.(2021)Iori, {de Lima}, Martello, Miyazawa and
  Monaci}]{Iori2021}
\bibinfo{author}{Iori, M.}, \bibinfo{author}{{de Lima}, V.L.},
  \bibinfo{author}{Martello, S.}, \bibinfo{author}{Miyazawa, F.K.},
  \bibinfo{author}{Monaci, M.}, \bibinfo{year}{2021}.
\newblock \bibinfo{title}{Exact solution techniques for two-dimensional cutting
  and packing}.
\newblock \bibinfo{journal}{European Journal of Operational Research}
  \bibinfo{volume}{289}, \bibinfo{pages}{399--415}.
\bibitem[{{Ministry of Health of Chile}(2021)}]{MHC2021}
\bibinfo{author}{{Ministry of Health of Chile}}, \bibinfo{year}{2021}.
\newblock \URLprefix \url{https://www.gob.cl/coronavirus/cifrasoficiales/}.
  \bibinfo{note}{accessed at July 21, 2021}.
\bibitem[{Pacheco and Laguna(2020)}]{Pacheco2020}
\bibinfo{author}{Pacheco, J.}, \bibinfo{author}{Laguna, M.},
  \bibinfo{year}{2020}.
\newblock \bibinfo{title}{{Vehicle routing for the urgent delivery of face
  shields during the {COVID‑19} pandemic}}.
\newblock \bibinfo{journal}{Journal of Heuristics} \bibinfo{volume}{26},
  \bibinfo{pages}{619--635}.
\bibitem[{Park et~al.(2021)Park, Kim, Kim, Lee and Giroux}]{Park2021}
\bibinfo{author}{Park, I.J.}, \bibinfo{author}{Kim, J.}, \bibinfo{author}{Kim,
  S.S.}, \bibinfo{author}{Lee, J.C.}, \bibinfo{author}{Giroux, M.},
  \bibinfo{year}{2021}.
\newblock \bibinfo{title}{{Impact of the COVID-19 pandemic on travelers’
  preference for crowded versus non-crowded options}}.
\newblock \bibinfo{journal}{Tourism Management} \bibinfo{volume}{87},
  \bibinfo{pages}{104398}.
\bibitem[{Roggeveen et~al.(2020)Roggeveen, Grewal and
  Schweiger}]{Roggeveen2020}
\bibinfo{author}{Roggeveen, A.L.}, \bibinfo{author}{Grewal, D.},
  \bibinfo{author}{Schweiger, E.B.}, \bibinfo{year}{2020}.
\newblock \bibinfo{title}{{The DAST Framework for Retail Atmospherics: The
  Impact of In- and Out-of-Store Retail Journey Touchpoints on the Customer
  Experience}}.
\newblock \bibinfo{journal}{Journal of Retailing} \bibinfo{volume}{96},
  \bibinfo{pages}{128--137}.
\newblock \bibinfo{note}{Understanding Retail Experiences and Customer Journey
  Management}.
\bibitem[{Sato et~al.(2013)Sato, de~Castro~Martins and de~Sales
  Guerra~Tsuzuki}]{Sato2013}
\bibinfo{author}{Sato, A.K.}, \bibinfo{author}{de~Castro~Martins, T.},
  \bibinfo{author}{de~Sales Guerra~Tsuzuki, M.}, \bibinfo{year}{2013}.
\newblock \bibinfo{title}{Placement heuristics for irregular packing to create
  layouts with exact placements for two moveable items}.
\newblock \bibinfo{journal}{IFAC Proceedings Volumes} \bibinfo{volume}{46},
  \bibinfo{pages}{384--389}.
\newblock \bibinfo{note}{11th IFAC Workshop on Intelligent Manufacturing
  Systems}.
\bibitem[{Seccia(2020)}]{Seccia2020}
\bibinfo{author}{Seccia, R.}, \bibinfo{year}{2020}.
\newblock \bibinfo{title}{{The Nurse Rostering Problem in {COVID-19} emergency
  scenario}}.
\newblock \URLprefix
  \url{http://www.optimization-online.org/DB\_FILE/2020/03/7712.pdf}.
\bibitem[{Silva et~al.(2016)Silva, Oliveira and Wäscher}]{Silva2016}
\bibinfo{author}{Silva, E.}, \bibinfo{author}{Oliveira, J.F.},
  \bibinfo{author}{Wäscher, G.}, \bibinfo{year}{2016}.
\newblock \bibinfo{title}{The pallet loading problem: {A} review of solution
  methods and computational experiments}.
\newblock \bibinfo{journal}{International Transactions in Operational Research}
  \bibinfo{volume}{23}, \bibinfo{pages}{147--172}.
\bibitem[{Somsen et~al.(2020)Somsen, van Rijn, Kooij, Bem and
  Bonn}]{somsen2020small}
\bibinfo{author}{Somsen, G.A.}, \bibinfo{author}{van Rijn, C.},
  \bibinfo{author}{Kooij, S.}, \bibinfo{author}{Bem, R.A.},
  \bibinfo{author}{Bonn, D.}, \bibinfo{year}{2020}.
\newblock \bibinfo{title}{Small droplet aerosols in poorly ventilated spaces
  and {SARS-CoV-2} transmission}.
\newblock \bibinfo{journal}{The Lancet. Respiratory Medicine} .
\bibitem[{Ugail et~al.(2021)Ugail, Aggarwal, Iglesias, Howard, Campuzano,
  Suárez, Maqsood, Aadil, Mehmood, Gleghorn, Taif, Kadry and
  Muhammad}]{Ugail2021}
\bibinfo{author}{Ugail, H.}, \bibinfo{author}{Aggarwal, R.},
  \bibinfo{author}{Iglesias, A.}, \bibinfo{author}{Howard, N.},
  \bibinfo{author}{Campuzano, A.}, \bibinfo{author}{Suárez, P.},
  \bibinfo{author}{Maqsood, M.}, \bibinfo{author}{Aadil, F.},
  \bibinfo{author}{Mehmood, I.}, \bibinfo{author}{Gleghorn, S.},
  \bibinfo{author}{Taif, K.}, \bibinfo{author}{Kadry, S.},
  \bibinfo{author}{Muhammad, K.}, \bibinfo{year}{2021}.
\newblock \bibinfo{title}{{Social distancing enhanced automated optimal design
  of physical spaces in the wake of the COVID-19 pandemic}}.
\newblock \bibinfo{journal}{Sustainable Cities and Society}
  \bibinfo{volume}{68}.
\bibitem[{Wang et~al.(2021)Wang, Yao and Martin}]{Wang2021}
\bibinfo{author}{Wang, D.}, \bibinfo{author}{Yao, J.}, \bibinfo{author}{Martin,
  B.A.}, \bibinfo{year}{2021}.
\newblock \bibinfo{title}{{The effects of crowdedness and safety measures on
  restaurant patronage choices and perceptions in the COVID-19 pandemic}}.
\newblock \bibinfo{journal}{International Journal of Hospitality Management}
  \bibinfo{volume}{95}, \bibinfo{pages}{102910}.
\bibitem[{Zucchi et~al.(2021)Zucchi, Iori and Subramanian}]{Zucchi2020}
\bibinfo{author}{Zucchi, G.}, \bibinfo{author}{Iori, M.},
  \bibinfo{author}{Subramanian, A.}, \bibinfo{year}{2021}.
\newblock \bibinfo{title}{{Personnel scheduling during {COVID-19} pandemic}}.
\newblock \bibinfo{journal}{Optimization Letters} \bibinfo{volume}{15},
  \bibinfo{pages}{1385--1396}.

\end{thebibliography}

\end{document}